\documentclass[showpacs,amsmath,amssymb,10pt]{revtex4}% Physical Review D
\allowdisplaybreaks

\usepackage{graphicx}
\usepackage{amssymb}
\usepackage{epstopdf}
\usepackage{epsfig}
\begin{document}

\def\aprge{\buildrel > \over {_{\sim}}}
\def\aprle{\buildrel < \over {_{\sim}}}

\def\etal{{\it et.~al.}}
\def\ie{{\it i.e.}}
\def\eg{{\it e.g.}}

\def\bwt{\begin{widetext}}
\def\ewt{\end{widetext}}
\def\be{\begin{equation}}
\def\ee{\end{equation}}
\def\bea{\begin{eqnarray}}
\def\eea{\end{eqnarray}}
\def\bean{\begin{eqnarray*}}
\def\eean{\end{eqnarray*}}
\def\bary{\begin{array}}
\def\eary{\end{array}}
\def\bi{\bibitem}
\def\bit{\begin{itemize}}
\def\eit{\end{itemize}}

\def\lan{\langle}
\def\ran{\rangle}
\def\lra{\leftrightarrow}
\def\la{\leftarrow}
\def\ra{\rightarrow}
\def\dash{\mbox{-}}
\def\ol{\overline}

\def\ub{\ol{u}}
\def\db{\ol{d}}
\def\sb{\ol{s}}
\def\cb{\ol{c}}

\def\re{\rm Re}
\def\im{\rm Im}

\def \b{{\cal B}}
\def \ca{{\cal A}}
\def \ko{K^0}
\def \ok{\overline{K}^0}
\def \s{\sqrt{2}}
\def \st{\sqrt{3}}
\def \sx{\sqrt{6}}
%\begin{document}
%\begin{large}

%\hspace{14.0cm}\hfill {\tt CAS-KITPC/ITP-086}\\

\title{{\bf Strong and radiative decays of $X(3872)$ as a hadronic molecule with a negative parity}}

\author{Masayasu Harada}
\author{Yong-Liang Ma}

%\email{^\dag ylma@itp.ac.cn}
\address{Department of Physics, Nagoya University, Nagoya, 464-8602, Japan.}
\date{\today}
%%%%%%%%%%%%%%%%%%%%%%%%%%%%%%%
\begin{abstract}
Properties of $X(3872)$ are studied by regarding it as a $DD^{\ast}$
hadronic molecule with $J^{PC} = 2^{-+}$ in the phenomenological
Lagrangian approach. We find that our model with about 97.6\%
isospin zero component explains the existing data nicely, for
example, the ratio $\mathcal{B}(X(3872) \to
J/\psi\pi^+\pi^-\pi^0)/\mathcal{B}(X(3872) \to J/\psi\pi^+\pi^-)$.
We predict the partial widths of the radiative decays of $X(3872)
\to \gamma J/\psi$, $\gamma \psi(2S) $ and the strong decays of
$X(3872) \to J/\psi \pi^+ \pi^-$, $J/\psi \pi^+\pi^-\pi^0$ as well
as $X(3872) \to \chi_{cJ}\pi^0$. Our analysis shows that the
measurement of the ratio $\mathcal{B}(X(3872) \to
\chi_{c0}\pi^0)/\mathcal{B}(X(3872) \to \chi_{c1}\pi^0)$ may signal
the nature of $X(3872)$.
\end{abstract}

\maketitle

\section{Introduction}

$X(3872)$ is one of the mysterious exotic states which was observed
by the Belle Collaboration in the exclusive $B^\pm \to K^\pm
\pi^+\pi^-J/\psi$ decays in 2003~\cite{Choi:2003ue}. Later, this
state was confirmed by the CDF~\cite{Acosta:2003zx}, the
D0~\cite{Abazov:2004kp} and the
BaBar~\cite{Aubert:2004ns,Aubert:2006aj,al.:2010jr} Collaborations.
The observation of its decay channels $\gamma
J/\psi$~\cite{Abe:2005ix,Aubert:2006aj} and $\gamma
\psi(2S)$~\cite{:2008rn} indicates that $X(3872)$ has a positive
charge parity. By analyzing the invariant $\pi^+\pi^-$ mass
distribution~\cite{Abulencia:2005zc} and the total angular
distribution~\cite{Abulencia:2006ma} of the $J/\psi \pi^+ \pi^-$
decay mode, the CDF Collaboration has narrowed its possible
spin-parity $(J^P)$ to be $1^+$ or $2^-$.

The $X(3872)$ is identified as an ``exotic" state because its
observed mass does not fit into the quark model predictions for the
charmonium
states~\cite{Eichten:1979ms,Buchmuller:1980su,Godfrey:1985xj} and,
the branching fraction of $X \to J\psi \omega \to J/\psi
\pi^+\pi^-\pi^0$ and $X \to J\psi \rho \to J/\psi \pi^+\pi^-$ is
almost equal. In the literature, $X(3872)$ has been interpreted as a
$c\bar{c}g$ hybrid state~\cite{Close:2003mb,Li:2004sta}, a
diquark-antidiquark state~\cite{Vijande:2004vt} , a $c\bar{c}$ and
meson cloudy mixing
state~\cite{Braaten:2005ai,Takeuchi:2007zz,Takizawa:2008zz,Dong:2008gb}
or a deuteron-like molecular state concerning its mass is
approximate to the threshold of $D^0D^{\ast
\,0}$~\cite{Tornqvist:2003na,Tornqvist:2004qy,Voloshin:2003nt,Wong:2003xk,Braaten:2003he,
Swanson:2003tb,Swanson:2004pp,Bugg:2004rk,Bugg:2004sh,Takeuchi:2007zza,Liu:2008tn,Dong:2009yp,Dong:2009uf,Fleming:2008yn}.
In addition, this state has also been studied with the coupling
channel
approach~\cite{Gamermann:2009fv,Gamermann:2009uq,Ortega:2010qq}. For
other studies of $X(3872)$ with $J^{PC} = 1^{++}$, see the recent
review Refs.~\cite{Drenska:2010kg,Biassoni:2010ew} and the
references therein.

Recently, the analysis of the $X(3872) \to J/\psi \omega \to J/\psi
\pi^+\pi^-\pi^0$ decay mode by the BaBar
Collaboration~\cite{al.:2010jr} indicated that $X(3872)$ favors a
negative parity, i.e., $J^{PC} = 2^{-+}$. Then, the radiative decays
of $X(3872) \to \gamma J/\psi$, $ \gamma\psi(2S)$ were studied by
assuming $X(3872)$ with $J^{PC} = 2^{-+}$ as the
$\eta_{c2}(1D)$~\cite{Jia:2010jn} and it was found that the data
cannot be interpreted. This indicates that $X(3872)$ may be an
exotic state.

In this paper, we study the decay properties of $X(3872)$ in the
effective Lagrangian approach by regarding it as a $DD^{\ast}$
molecule with $J^{PC} = 2^{-+}$. We use the method which was
proposed by one of us in the previous
work~\cite{Faessler:2007gv,Faessler:2007us,Faessler:2008vc} and was
applied to study the properties of $X(3872)$ with quantum numbers
$J^{PC} = 1^{++}$~\cite{Dong:2008gb,Dong:2009uf,Dong:2009yp} and
other exotic states~\cite{Ma:2008hc,Ma:2009sm,Ma:2010xx}. The
composite state $X(3872)$ is defined via the compositeness condition
$Z = 0$ which was originally used in the study of
deuteron~\cite{Weinberg:1962hj,Salam:1962ap} which is a bound state
of proton and neutron with $Z = 0$ as the wave function
renormalization constant of the composite particle.

We write the wave function of $X(3872)$ as
\begin{eqnarray}
|X(3872)\rangle = \frac{\cos\theta}{\sqrt{2}}| D^{0}\bar{D}^{\ast \,
0}\rangle + \frac{\sin\theta}{\sqrt{2}}| D^{+}D^{\ast \, -}\rangle +
{\rm C.c.},\label{eq:defmix}
\end{eqnarray}
which is in terms of charge eigenstate. Or equivalently, one may
regard the observed $X(3872)$ as a mixing state of $I = 0$ and $I =
1$ states, i.e.,
\begin{eqnarray}
|X(3872)\rangle = \cos\phi|X(3872)\rangle_{I=0} +
\sin\phi|X(3872)\rangle_{I=1}, \label{eq:mixiso}
\end{eqnarray}
with $\cos\theta = (\cos\phi + \sin\phi)/\sqrt{2}$ and $\sin\theta =
(\cos\phi - \sin\phi)/\sqrt{2}$ and
\begin{eqnarray}
|X(3872)\rangle_{I=0} = \frac{1}{2}\Big(| D^{0}\bar{D}^{\ast \,
0}\rangle + | D^{+}D^{\ast \, -}\rangle \Big) + {\rm C.c.},
\;\;\;\;\;\;\; |X(3872)\rangle_{I=1} = \frac{1}{2}\Big(|
D^{0}\bar{D}^{\ast \, 0}\rangle - | D^{+}D^{\ast \, -}\rangle \Big)
+ {\rm C.c.}.
\end{eqnarray}

We find, using the two parameters, the mixing angle $\phi$ between
the isospin singlet and triplet components introduced in
Eq.~(\ref{eq:mixiso}) and the size parameter $\Lambda_X$ introduced
in the following to illustrate the distribution of the constituents
in the molecule, we can explain the date quite well. From the
magnitude of the mixing angle $\phi$ which was fitted from the data,
we see that, in the wave function of $X(3872)$, the isospin singlet
component is dominant which is consistent with the analysis from the
phase space. That is, in case of $X(3872)$ as a tensor meson, the
phase space of the decay $X(3872) \to J/\psi \pi^+\pi^-\pi^0$ is
about one percent of that of the decay $X(3872) \to J/\psi
\pi^+\pi^-$ so only few percent of isospin one component in
$X(3872)$ can accommodate the large isospin violating partial width.
We also explicitly calculated other decay widths of $X(3872) \to
\gamma J/\psi$, $\gamma \psi(2S)$ and $X(3872) \to \chi_{cJ}\pi^0$
($J = 0, 1, 2$) using these two fixed parameters. Comparing our
results with the more precise data observed in the future, one can
determine the structure of $X(3872)$. For example, if the future
data are not consistent with our present results, there may be
$J/\psi\omega$, $J/\psi\rho$ or/and $c\bar{c}$ components in the
$X(3872)$ wave function.

This paper is organized as follows: In Sec.~\ref{sec:theoretical},
we propose the theoretical framework applied in our calculation. In
Sec.~\ref{sec:decay} we give the formalism for the strong and
radiative decays of $X(3872)$ explicitly. Sec.~{\ref{sec:numerical}}
is our numerical results and discussions and the last section is our
conclusion.

\section{The theoretical approach}

\label{sec:theoretical}

As was mentioned above, we regard $X(3872)$ as a $D D^{\ast}$ bound
state with $J^{PC} = 2^{-+}$ with the explicit wave function
(\ref{eq:defmix}). Following Ref.~\cite{Bellucci:1994eb}, one has
the free Lagrangian for tensor field as
\begin{eqnarray}
{\cal L}_{\rm free}^{T} & = & -\frac{1}{2} X_{\mu\nu}
D^{\mu\nu;\lambda\sigma} X_{\lambda\sigma} ,
\end{eqnarray}
where the symmetric tensor $X_{\mu\nu} = X_{\nu\mu}$ denotes the
$J^{PC} = 2^{-+}$ field for $X(3872)$ and
\begin{eqnarray}
D^{\mu\nu;\lambda\sigma} & = & (\Box +
m_{_X}^2)\Big\{\frac{1}{2}(g^{\mu\lambda}g^{\nu\sigma}+g^{\nu\lambda}g^{\mu\sigma})-g^{\mu\nu}g^{\lambda\sigma}\Big\}
\nonumber\\
& & +g^{\lambda\sigma}\partial^\mu\partial^\nu +
g^{\mu\nu}\partial^\lambda\partial^\sigma - \frac{1}{2}\Big[ \Big(
g^{\nu\sigma}\partial^\mu\partial^\lambda +
g^{\nu\lambda}\partial^\mu\partial^\sigma \Big) +(\mu
\leftrightarrow \nu) \Big] ,
\end{eqnarray}
with $m_{_X}$ being the mass for $X(3872)$. The propagator for
$X_{\mu\nu}(3872)$ is obtained as
\begin{eqnarray}
G_{\mu\nu;\lambda\sigma}(x) & = & \int\frac{d^4p}{(2\pi)^4}\frac{i}{p^2 - m_{_X}^2-i\epsilon}P_{\mu\nu;\lambda\sigma} e^{-ip\cdot x}  \; ,\nonumber\\
P_{\mu\nu;\lambda\sigma} & = &
\frac{1}{2}(P_{\mu\lambda}P_{\nu\sigma}+P_{\mu\sigma}P_{\nu\lambda})
- \frac{1}{3}P_{\mu\nu}P_{\lambda\sigma}  \, , \nonumber\\
P_{\mu\nu} & = & - g_{\mu\nu} + \frac{p_\mu p_\nu}{m_{_X}^2}  \; .
% D^{\mu\nu;\lambda\sigma}G_{\lambda\sigma}^{\,\,\,\,\,\, \alpha\beta}
% & = & -i\frac{1}{2}(g^{\mu\alpha}g^{\nu\beta} +
% g^{\nu\alpha}g^{\mu\beta})\delta^{(4)}(x) .
\end{eqnarray}

Taking into account the quantum numbers of $X(3872)$ and its
constituents $DD^{\ast}$, one can write down the effective
Lagrangian describing the interaction between $X(3872)$ and its
constituents as
\begin{eqnarray}
{\cal L}_{\rm X} & = & \frac{i}{\sqrt{2}}X^{\mu\nu}(x)\int dx_1
dx_2 \Phi_X((x_1-x_2)^2)\delta(x-\omega_v x_1 - \omega_px_2)\label{effelcomp}\\
& & \times \Big\{g_{_X}^{N}\Big[ C_{\mu\nu}^{N}(x_1,x_2) +
C_{\nu\mu}^{N}(x_1,x_2) -
\frac{1}{4}g_{\mu\nu}C_{\alpha}^{N;\alpha}(x_1,x_2)\Big] +
g_{_X}^{C}\Big[ C_{\mu\nu}^{C}(x_1,x_2) + C_{\nu\mu}^{C}(x_1,x_2) -
\frac{1}{4}g_{\mu\nu}C_{\alpha}^{C;\alpha}(x_1,x_2)\Big]
\Big\},\nonumber
\end{eqnarray}
where $g_{_X}^{N}$ is the effective coupling constant for the
interaction between $X(3872)$ and its neutral constituents $(D^{0}
\bar{D}^{\ast \, 0} + \bar{D}^{0} D^{\ast \, 0})$ while $g_{_X}^{C}$
is for the interaction between $X(3872)$ and its charged
constituents $(D^{+} D^{\ast \, -} + D^{-} D^{\ast \, +})$.
$\omega_v$ and $\omega_p$ are mass ratios with definitions
\begin{eqnarray}
\omega_v & = & \frac{m_{_{D^{\ast}}}}{m_{_{D^{\ast}}} + m_{_{D}}} \;
, \,\,\,\, \omega_p = \frac{m_{_{D}}}{m_{_{D^{\ast}}} + m_{_{D}}} ,
\end{eqnarray}
where $m_{_D}$ and $m_{_{D^{\ast}}}$ are masses of the constituents
pseudoscalar meson $D$ and vector meson $D^{\ast}$, respectively. In
the Lagrangian (\ref{effelcomp}), to illustrate the finite size of
the molecule, the function $\Phi_X((x_1-x_2)^2)$ with the Fourier
transform
\begin{eqnarray}
\Phi_X(y^2) & = &
\int\frac{d^4p}{(2\pi)^4}\tilde{\Phi}_X(p^2)e^{-ip\cdot y} ,
\end{eqnarray}
has been introduced. And, for simplicity, we have defined the tensor
$C_{\mu\nu}$ as a function of the constituents with the explicit
form in terms of the neutral constituents as
\begin{eqnarray}
C_{\mu\nu}^{N}(x_1,x_2) & = & \bar{D}_{\mu}^{\ast \,
0}(x_1)\partial_\nu D^{0}(x_2) + D_{\nu}^{\ast \,
0}(x_1)\partial_\mu \bar{D}^{0}(x_2) . \nonumber
\end{eqnarray}
Substituting the fields $D^{0}$ and $D^{\ast \, 0}$ with the
corresponding charged ones, one can get the explicit form of
$C_{\mu\nu}^{C}$.

\begin{figure}[htbp]
\begin{center}
\includegraphics[scale=0.7]{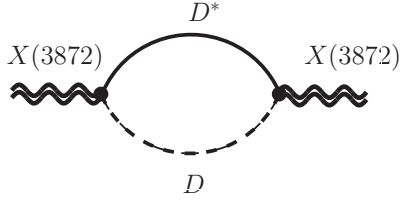}
\end{center}
\caption[The mass operator for $X(3872)$.]{%
The mass operator for $X(3872)$. } \label{fig:MassOperator}
\end{figure}

Relation between the mixing angle $\theta$ defined in
Eq.~(\ref{eq:defmix}) and the coupling constant
$g_{_X}^{N}(g_{_X}^{C})$ can be yielded with the help of the
compositeness condition $Z_X = 0$ with $Z_X$ as the wave function
renormalization constant of $X(3872)$. $Z_X$ is defined as the
residual of $X(3872)$ propagator, i.e.,
\begin{eqnarray}
Z_X & = & 1 -
g_{_X}^2\frac{d}{dp^2}\Sigma_{_X}(p^2)\Big|_{p^2=m_{_X}^2} ,
\end{eqnarray}
where $g_{_X}^2\Sigma_{_X}(p^2)$ relates to the mass operator via
the relation
\begin{eqnarray}
\Pi_{_X}^{\mu\nu;\alpha\beta}(p^2) & = &
\frac{1}{2}(g^{\mu\alpha}g^{\nu\beta} +
g^{\mu\beta}g^{\nu\alpha})g_{_X}^2\Sigma_{_X}(p^2) + \cdots ,
\end{eqnarray}
with $``\cdots"$ denoting terms do not contribute to the mass
renormalization of $X(3872)$. The mass operator of $X(3872)$ is
illustrated in Fig.~\ref{fig:MassOperator}.
Explicitly, we have the following relations between the mixing angle
$\theta$ and $g_{_X}^{N}$ and $g_{_X}^{C}$
\begin{eqnarray}
\cos\theta & = &
g_{_X}^{N}\frac{d}{dp^2}\Sigma_{_X}^{N}(p^2)\Big|_{p^2 =
m_{_X}^2},\;\;\;\;\; \cos\theta =
g_{_X}^{C}\frac{d}{dp^2}\Sigma_{_X}^{C}(p^2)\Big|_{p^2 = m_{_X}^2},
\end{eqnarray}
where $\Sigma_{_X}^{N}$ corresponds to the case with neutral
constituents but $\Sigma_{_X}^{C}$ corresponds to the case with
charged constituents.

The mass operator can be calculated from the Feynman diagram
depicted in Fig.~\ref{fig:MassOperator}. In our calculation we
applied the Gaussian form of $\tilde{\Phi}_X(p^2)$, i.e.,
\begin{eqnarray}
\tilde{\Phi}_X(p^2) & = & \exp(p^2/\Lambda_{_X}^2) ,
\label{gaussian}
\end{eqnarray}
with the size parameter $\Lambda_{_X}$ introduced to parameterize
the distribution of the constituents inside the molecule. In the
following calculation, we will take $\Lambda_{_X}$ as a free
parameter and fix it from the data.

Because of the uncertainty of the $X(3872)$ mass measurement, we
express the mass of $X(3872)$ in terms of the binding energy as
\begin{eqnarray}
m_{_X} & = & m_{D^{\ast \, 0}} + m_{D^0} - \Delta E ,
\label{eq:bindenergy}
\end{eqnarray}
where $\Delta E > 0$ is the binding energy and $m_{D^{\ast \, 0}} =
2006.97~$MeV and $m_{D^0} = 1864.84~$MeV~\cite{Amsler:2008zz}. In
the following calculation, we take $\Delta E = 0.5, 1.0$ and
$1.5~$MeV.

%After standard loop calculation, the numerical result of $g_{_X}$ in
%the typical region $2.0~{\rm GeV} \leq\Lambda_{_X} \leq 3.0~{\rm
%GeV}$ can be yielded as
%\begin{eqnarray}
%g_{_X} = 19.71_{-0.13}^{+0.12} - 22.71_{-0.14}^{+0.13} ,
%\end{eqnarray}
%where the ambiguity is due to the change of $\Delta E$ and $g_{_X}$
%decreases with the increasing of $\Lambda_{_X}$. In
%Fig.~\ref{line:gx} we plot the $\Lambda_{_X}$ dependence of $g_{_X}$
%in the region $1.0~{\rm GeV} \leq\Lambda_{_X} \leq 3.0~{\rm GeV}$.
%\begin{figure}[htbp]
%\begin{center}
%\includegraphics[scale=0.8]{linemass.eps}
%\end{center}
%\caption[The $\Lambda_{_X}$ dependence of the coupling constant $g_{_X}$.]{%
%The $\Lambda_{_X}$ dependence of the coupling constant $g_{_X}$. }
%\label{line:gx}
%\end{figure}

%\subsection{Effective Lagrangian for Strong and Electromagnetic Decays of $X(4350)$}

To calculate the strong decays, we will use the effective Lagrangian
including the following terms
\begin{eqnarray}
{\cal L}_{D^{\ast}D^{\ast}V} & = & i g_{D^{\ast}D^{\ast}\psi_n}[\bar{D}_{\mu}^{\ast}(D_{\nu}^{\ast}\partial^{^{^{\hspace{-0.2cm}\leftrightarrow}}\mu} \psi_{n}^{\nu}) + D_{\mu}^{\ast}(\psi_{n;\, \nu}\partial^{^{^{\hspace{-0.2cm}\leftrightarrow}}\mu} \bar{D}^{\ast\nu} ) + \psi_{n;\, \mu} (\bar{D}_{\nu}^{\ast}\partial^{^{^{\hspace{-0.2cm}\leftrightarrow}}\mu} D^{\ast\nu})] \nonumber\\
& & + i g_{D^{\ast}D^{\ast}\omega}[\bar{D}_{\mu}^{\ast}(D_{\nu}^{\ast}\partial^{^{^{\hspace{-0.2cm}\leftrightarrow}}\mu} \omega^{\nu}) + D_{\mu}^{\ast}(\omega_{\nu}\partial^{^{^{\hspace{-0.2cm}\leftrightarrow}}\mu} \bar{D}^{\ast\nu} ) + \omega_{\mu} (\bar{D}_{\nu}^{\ast}\partial^{^{^{\hspace{-0.2cm}\leftrightarrow}}\mu} D^{\ast\nu})] \nonumber\\
& & + i g_{D^{\ast}D^{\ast}\rho}[\bar{D}_{\mu}^{\ast}(D_{\nu}^{\ast}\partial^{^{^{\hspace{-0.2cm}\leftrightarrow}}\mu} \vec{\tau}\cdot\vec{\rho}^{\;\nu}) + D_{\mu}^{\ast}(\vec{\tau} \cdot \vec{\rho}_{\nu}\partial^{^{^{\hspace{-0.2cm}\leftrightarrow}}\mu} \bar{D}^{\ast\nu} ) + \vec{\tau}\cdot \vec{\rho}_{\mu} (\bar{D}_{\nu}^{\ast}\partial^{^{^{\hspace{-0.2cm}\leftrightarrow}}\mu} D^{\ast\nu})],\nonumber\\
%%%%%%%%%%%%%%%%%%%%%%%%%%%%%%%%%%%%%%%%%%%%%%%%%%%%%%%%%%%%%%%%%%%%%%%%%%%%%%%%%%%%%%%%%%%%%%%%%%%%%%%%%%%%%%%%%%
{\cal L}_{DDV} & = & - i g_{DD\psi_n} \psi_{n;\, \mu} (\bar{D}\partial^{^{^{\hspace{-0.2cm}\leftrightarrow}}\mu} D) - i g_{DD\omega} \omega_{\mu} (\bar{D}\partial^{^{^{\hspace{-0.2cm}\leftrightarrow}}\mu} D) - i g_{DD\rho} \vec{\rho}_{\mu}\cdot(\bar{D}\vec{\tau}\partial^{^{^{\hspace{-0.2cm}\leftrightarrow}}\mu} D) ,\nonumber\\
%%%%%%%%%%%%%%%%%%%%%%%%%%%%%%%%%%%%%%%%%%%%%%%%%%%%%%%%%%%%%%%%%%%%%%%%%%%%%%%%%%%%%%%%%%%%%%%%%%%%%%%%%%%%%%%%%%
{\cal L}_{D^{\ast}DV} & = &
g_{D^{\ast}D\psi_n}\epsilon^{\mu\nu\alpha\beta} \, \psi_{n;\,\mu\nu}
\bar{D}_{\alpha\beta}^{\ast} D \nonumber\\
& & + g_{D^{\ast}D\omega}\epsilon^{\mu\nu\alpha\beta} \,
\omega_{\mu\nu} \bar{D}_{\alpha\beta}^{\ast} D +
g_{D^{\ast}D\rho}\epsilon^{\mu\nu\alpha\beta} \, \rho_{\mu\nu} \cdot
\bar{D}_{\alpha\beta}^{\ast}\vec{\tau} D + {\rm
H.c.} ,\nonumber\\
%%%%%%%%%%%%%%%%%%%%%%%%%%%%%%%%%%%%%%%%%%%%%%%%%%%%%%%%%%%%%%%%%%%%%%%%%%%%%%%%%%%%%%%%%%%%%%%%%%%%%%%%%%%%%%%%%%
{\cal L}_{D^{\ast}D\pi} & = &
\frac{ig_{D^{\ast}D\pi}}{2\sqrt{2}}\bar{D}^{\ast}_\mu
(\vec{\tau}\cdot \vec{\pi}
\partial^{^{^{\hspace{-0.2cm}\leftrightarrow}}\mu} D ) + {\rm
H.c.} , \nonumber\\
%%%%%%%%%%%%%%%%%%%%%%%%%%%%%%%%%%%%%%%%%%%%%%%%%%%%%%%%%%%%%%%%%%%%%%%%%%%%%%%%%%%%%%%%%%%%%%%%%%%%%%%%%%%%%%%%%%
{\cal L}_{D^{\ast}D^{\ast}\pi} & = &
\frac{g_{D^{\ast}D^{\ast}\pi}}{2\sqrt{2}}\epsilon^{\mu\nu\alpha\beta}\bar{D}^{\ast}_{\mu\nu}\vec{\tau}\cdot \vec{\pi}D^{\ast}_{\alpha\beta} , \nonumber\\
%%%%%%%%%%%%%%%%%%%%%%%%%%%%%%%%%%%%%%%%%%%%%%%%%%%%%%%%%%%%%%%%%%%%%%%%%%%%%%%%%%%%%%%%%%%%%%%%%%%%%%%%%%%%%%%%%%
{\cal L}_{\chi_{c0}DD} & = & g_{\chi_{c0}DD}\chi_{c0}\bar{D} D , \nonumber\\
%%%%%%%%%%%%%%%%%%%%%%%%%%%%%%%%%%%%%%%%%%%%%%%%%%%%%%%%%%%%%%%%%%%%%%%%%%%%%%%%%%%%%%%%%%%%%%%%%%%%%%%%%%%%%%%%%%
{\cal L}_{\chi_{c0}D^{\ast}D^{\ast}} & = & g_{\chi_{c0}D^{\ast}D^{\ast}}\chi_{c0}\bar{D}^{\ast}_\mu D^{\ast\mu} , \nonumber\\
%%%%%%%%%%%%%%%%%%%%%%%%%%%%%%%%%%%%%%%%%%%%%%%%%%%%%%%%%%%%%%%%%%%%%%%%%%%%%%%%%%%%%%%%%%%%%%%%%%%%%%%%%%%%%%%%%%
{\cal L}_{\chi_{c1}D^{\ast}D} & = & ig_{\chi_{c1}D^{\ast}D}\chi_{c1;\,\mu}\bar{D} D^{\ast\mu} + {\rm H.c.}, \nonumber\\
%%%%%%%%%%%%%%%%%%%%%%%%%%%%%%%%%%%%%%%%%%%%%%%%%%%%%%%%%%%%%%%%%%%%%%%%%%%%%%%%%%%%%%%%%%%%%%%%%%%%%%%%%%%%%%%%%%
{\cal L}_{\chi_{c2}D^{\ast}D^{\ast}} & =
&g_{\chi_{c2}D^{\ast}D^{\ast}}\chi_{c2;\,\mu\nu}
[\bar{D}^{\ast\mu}D^{\ast\nu} +
\bar{D}^{\ast\nu}D^{\ast\mu}],\label{Eq.lagstr}
\end{eqnarray}
where $\psi_n$ denotes $J/\psi$ and $\psi(2S)$ vector mesons,
$\vec{\pi}$ and $\vec{\rho}$ stand for the pion and rho meson
triplets respectively and  $D = (D^0,D^+)^T$ and $D^{\ast} =
(D^{\ast \, 0},D^{\ast \, +})^T$ are the pseudoscalar and the vector
charmed meson doublets respectively. $V_{\mu\nu}~(V = \psi_n,
D^{\ast}, \omega, \rho)$ is defined as $V_{\mu\nu}  =
\partial_\mu V_\nu -\partial_\nu V_\mu$. The definition of the right-left derivative
$A\partial^{^{^{\hspace{-0.2cm}\leftrightarrow}}}B$ is
$A\partial^{^{^{\hspace{-0.2cm}\leftrightarrow}}}B \equiv A
(\partial B) - (\partial A)B$. In Ref.~\cite{Lin:1999ad}, the
coupling constants $g_{D^{\ast}D^\ast V}$ and $g_{DD V}$ were
determined with the help of the VMD, i.e.,
\begin{eqnarray}
g_{D^{\ast}D^\ast \rho} & = & g_{DD \rho} = -2.52 \; ,\;\;\;
g_{D^{\ast}D^\ast \omega} = g_{DD \omega} = -2.84 \; , \;\;\;
g_{D^{\ast}D^\ast \psi} = g_{DD \psi} = 7.64 \; .
\label{eq:couplingddv}
\end{eqnarray}
Using the VMD and the partial widths $\Gamma(J/\psi \to e^+e^-) =
5.55~{\rm KeV}$ and $\Gamma(\psi(2S) \to e^+e^-)=2.38~{\rm
KeV}$~\cite{Amsler:2008zz}, we have
\begin{eqnarray}
\frac{g_{D^{\ast}D^\ast\psi^\prime}}{g_{D^{\ast}D^\ast \psi}} =
\frac{g_{DD \psi^\prime}}{g_{DD \psi}} =
\frac{m_{_{\psi^\prime}}}{m_\psi}\sqrt{\frac{\Gamma(\psi \to
e^+e^-)}{\Gamma(\psi^\prime \to e^+e^-)}} = 1.82,
\end{eqnarray}
which leads to $g_{D^{\ast}D^\ast\psi^\prime} = g_{DD \psi^\prime} =
13.90$.

From the HHChPT including vector mesons within the hidden local
symmetry
method~\cite{Kitazawa:1993bk,Harada:2003kt,Casalbuoni:1996pg} (see
Ref.~\cite{Jain:1994rb} for an equivalent approach), one can yield
the following relation for the coupling
constants~\cite{Casalbuoni:1996pg}
\begin{eqnarray}
g_{D^{\ast}D\omega} & = &g_{D^{\ast}D\rho} =
\frac{\lambda}{2}g\sqrt{\frac{m_{_D}}{m_{_{D^\ast}}}},
\end{eqnarray}
where $g$ is the universal coupling constant introduced in the
hidden local symmetry method with $g = 5.8 \pm 0.91$. And, $\lambda$
gives the coupling of the light vector meson with the heavy states
and as analyzed in Ref.~\cite{Casalbuoni:1996pg}, $\lambda=
-0.41~$GeV$^{-1}$. Finally, we get $g_{D^{\ast}D\omega} =
 g_{D^{\ast}D\rho} = -1.23~$GeV$^{-1}$.

As was discussed in Ref.~\cite{Faessler:2007gv}, we fix the coupling
constant $g_{D^{\ast} D \pi}$ from the partial width for the decay
of $D^{\ast} \to D \pi$ which leads to $g_{D^{\ast} D \pi} = 17.9$.
$g_{D^{\ast} D^{\ast} \pi}$ is related to $g_{D^{\ast} D \pi}$ via
HHChPT~\cite{Wise:1992hn}, i.e.,
\begin{eqnarray}
g_{D^{\ast} D^{\ast} \pi} & = & \frac{g_{D^{\ast} D
\pi}}{4\sqrt{m_{_D}m_{_{D^{\ast}}}}} = 2.31{~\rm GeV}^{-1} \,.
\label{eq:gdstdstpidstdpi}
\end{eqnarray}
In Ref.~\cite{Colangelo:2003sa}, the HHChPT for the interaction of
charmonium with open charm mesons was constructed. From this theory,
one can yield the following relations for the coupling constants
\begin{eqnarray}
g_{D^{\ast}D\psi_n} & = &
-\frac{1}{2}g_{DD\psi_n}\frac{\sqrt{m_{_{D^{\ast}}}}}{m_{\psi_n}\sqrt{m_{D}}},
%= -0.17{~\rm GeV}^{-1} \, ,
\\
g_{\chi_{c0}DD} & = & 3 \frac{m_{D}}{m_{D^\ast}} g_{\chi_{c0}D^\ast
D^\ast} = - 2\sqrt{3} g_1 m_D \sqrt{ m_{\chi_{c0}}} \,, \label{eq:gchic0dd}\\
g_{\chi_{c1}D^\ast D} & = & \sqrt{2}g_1 \sqrt{m_{\chi_{c1}} m_{D} m_{D^\ast}} \,, \label{eq:gchic1dstd}\\
g_{\chi_{c2}D^\ast D^{\ast}} & = & 2 g_1 m_{D^\ast}
\sqrt{m_{\chi_{c2}}},
\end{eqnarray}
with the expression for $g_1$ as
\begin{eqnarray}
g_1 = - \sqrt{\frac{m_{\chi_{c0}}}{3}} \frac{1}{f_{\chi_{c0}}} \,,
\end{eqnarray}
where $f_{\chi_{c0}}$ is defined via the relation $\langle 0 | \bar
c c | \chi_{c0}(p) \rangle = f_{\chi_{c0}} m_{\chi_{c0}}$ and the
QCD sum rules yield $f_{\chi_{c0}} = 510$
MeV~\cite{Colangelo:2002mj}. So that we have $g_1 = - 2.09$
GeV$^{-1/2}$.

The electromagnetic interaction Lagrangian applied in our
calculation can be decomposed as
\begin{eqnarray}
{\cal L}_{\rm em} &
= & {\cal L}_{\rm em}^{\rm NL} + {\cal L}_{\rm kin}^{\rm gauge} +
{\cal L}_{D^{\ast}D\psi_n\gamma} + {\cal L}_{D^{\ast}D\gamma}.
\label{eq.lagem}
\end{eqnarray}

The effective Lagrangian ${\cal L}_{\rm em}^{\rm NL}$ is yielded by
substituting $C_{\mu\nu}^{C}$ in Eq.~(\ref{effelcomp}) with
$C_{\mu\nu}^{C;\,\rm gauge}$ with the explicit form
\begin{eqnarray}
C_{\mu\nu}^{\rm C; \, gauge}(x_1,x_2) & = & e^{ieI(x_1,x_2;P)}
D_{\mu}^{\ast \, -}(x_1)[\partial_\nu - ieA_\nu (x_2)] D^{+}(x_2) +
e^{-ieI(x_1,x_2;P)} D_{\nu}^{\ast \, +}(x_1)[\partial_\mu + ieA_\mu
(x_2)] D^{-}(x_2), \nonumber
\end{eqnarray}
where the Wilson's line $I(x,y,P)$ is defined as
\begin{eqnarray}
I(x,y;P) & = & \int_y^x dz_\mu A^\mu(z) \; .
\end{eqnarray}
In the following calculation, the nonlocal vertex with one photon is
necessary. This vertex comes from two sources: One is from the
covariant derivative and the other one is from the expansion of the
Wilson's line. One can easily derive the Feynman rule for the
nonlocal vertex with one-photon coming from the covariant
derivative. But to derive the Feynman rule for the photon from
Wilson's line, one may use the path-independent prescription
suggested in~\cite{Mandelstam:1962us,Holdom:1990iq,Terning:1991yt}.

${\cal L}_{\rm em}^{\rm gauge}$ is from the gauged kinetic terms of
the charged constituents $D^{\ast \, +}$ and $D^{+}$, i.e.,
\begin{eqnarray}
{\cal L}_{\rm kin}^{\rm gauge} & = & ie A_\mu
(D^{-}\partial^{^{^{\hspace{-0.2cm}\leftrightarrow}}\mu} D^{+}) +
ieA_\mu [ - D_{\alpha}^{\ast \,
-}\partial^{^{^{\hspace{-0.2cm}\leftrightarrow}}\mu} D^{\ast \, + \,
\alpha} + \frac{1}{2}D_{\alpha}^{\ast \,
-}\partial^{^{^{\hspace{-0.2cm}\leftrightarrow}}\alpha} D^{\ast \, +
\, \mu} + \frac{1}{2}D^{\ast \, - \,
\mu}\partial^{^{^{\hspace{-0.2cm}\leftrightarrow}}}_\alpha D^{\ast
\, + \, \alpha}] \; .
\end{eqnarray}
And ${\cal L}_{D^{\ast}D\psi_n\gamma}$ is obtained by gauging the
derivative coupling of the $D^{\ast}D\psi_n$ vertex in
Eq.~(\ref{Eq.lagstr})
\begin{eqnarray}
{\cal L}_{D^{\ast}D\psi_n\gamma} & = &
g_{D^{\ast}D\psi_n}\epsilon^{\mu\nu\alpha\beta} \, \psi_{n;\,\mu\nu}
\hat{D}_{\alpha\beta}^{\ast \, -} D^+ + {\rm H.c.} ,
\end{eqnarray}
with $\hat{D}_{\alpha\beta}^{\ast \, -}$ defined via the covariant
derivative as $\hat{D}_{\alpha\beta}^{\ast \, -} = (\partial +
ieA)_\alpha D_\beta^{\ast \, -} - (\partial + ieA)_\beta
D_\alpha^{\ast \, -}$.

The effective Lagrangian ${\cal L}_{D^{\ast}D\gamma}$ can be
generally expressed in the form
\begin{eqnarray}
{\cal L}_{D^{\ast}D\gamma} & = & e g_{D^{\ast
\,+}D^+\gamma}\epsilon^{\mu\nu\alpha\beta}D^+D^{\ast \,
-}_{\mu\nu}F_{\alpha\beta} -  e g_{D^{\ast
\,0}D^0\gamma}\epsilon^{\mu\nu\alpha\beta}D^0\bar{D}^{\ast \,
0}_{\mu\nu}F_{\alpha\beta}\label{eq:dstdgamma}.
\end{eqnarray}
From the partial width for the decay $D^{\ast} \to D \gamma$ one can
determine the coupling constant $g_{D^{\ast}D\gamma}$. Using the
data $\Gamma(D^{\ast \, +} \to D^+\gamma) = 1.54~$KeV, we have
$g_{D^{\ast \,+}D^+\gamma} \simeq 0.124~$GeV$^{-1}$. With respect to
the isospin invariance, the decay $D^{\ast \,0} \to D^0 \pi^0$ can
be related to $D^{\ast \,+} \to D^0 \pi^+$~\cite{Dong:2008gb}. And
considering the fact that the branching ratios for the decays
$D^{\ast \,0} \to D^0 \gamma$ and $D^{\ast \,0} \to D^0 \pi^0$ are
well measured~\cite{Amsler:2008zz}, one can yield the coupling
constant $g_{D^{\ast \,0}D\gamma} \simeq 0.51~$GeV$^{-1}$. The
relative sign between these two terms are fixed as in
Ref.~\cite{Manohar:2000dt}.
%Since to our knowledge, we do not no how to determine the relative
%sigh between ${\cal L}_{D^{\ast}D\gamma}$ and other terms, we leave
%this as an ambiguity of our calculation and consider the two cases
%as indicated by Eq.~(\ref{eq.lagem}).

\section{Strong and Radiative decays of $X(3872)$}

\label{sec:decay}

\subsection{Radiative $X(3872) \to \gamma J/\psi $ and $X(3872) \to \gamma \psi(2S)$ decays}

In this part, we study the radiative decays of $X(3872)$. Taking
into account the quantum numbers of the initial and final states,
one can write the most general form of the gauge invariant matrix
element for the electromagnetic decay of $X(3872) \to \gamma(q_1)
V(q_2)$ as
\begin{eqnarray}
iM_{XV\gamma} & = & ieM_{XV\gamma}^{\mu\nu\alpha\beta}\epsilon_{\mu\nu}(p)\epsilon_\alpha^{\ast}(q_1)\epsilon_\beta^{\ast}(q_2) \nonumber\\
& = & ie \Big\{\alpha_{XV\gamma}^{(1)} \Big[\epsilon^{\mu\sigma\alpha\rho}g^{\nu\beta}p_\sigma q_\rho + (\mu \leftrightarrow \nu) \Big] + \alpha_{XV\gamma}^{(2)} \Big[\epsilon^{\mu\sigma\rho\beta}p_\sigma q_\rho[g^{\nu\alpha} - \frac{(p+q)^\nu q^\alpha}{q\cdot(p+q)}] + (\mu \leftrightarrow \nu) \Big]\nonumber\\
& & + \alpha_{XV\gamma}^{(3)} \Big[\epsilon^{\mu\sigma\alpha\beta}(p+q)_\sigma q^\nu + (\mu \leftrightarrow \nu) \Big] + \alpha_{XV\gamma}^{(4)} \Big[\epsilon^{\mu\sigma\alpha\rho}p_\sigma q_\rho q^\nu q^\beta + (\mu \leftrightarrow \nu) \Big] \nonumber\\
& & + \alpha_{XV\gamma}^{(5)}
\Big[\epsilon^{\sigma\rho\alpha\beta}p_\sigma q_\rho q^\mu q^\nu
\Big]\Big\}\epsilon_{\mu\nu}(p)\epsilon_{\alpha}^{\ast}(q_1)\epsilon_{\beta}^{\ast}(q_2),
\label{matrixvg}
\end{eqnarray}
where $\epsilon_{\mu\nu}$ is the polarization tensor for the tensor
meson $X(3872)$ satisfying the transverseness condition $p^\mu
\epsilon_{\mu\nu}(p) = 0$ and the traceless condition
$\epsilon_\mu^\mu(p) = 0$. $q$ is defined as $q = q_1-q_2$ and $q_i$
satisfies the transverseness condition $q_i\cdot \epsilon(q_i) = 0$.

%It should be noted that, the expression (\ref{matrixvg}) can be
%induced by imposing the gauge invariance on Eq.~(\ref{matrixvv})
%given in the following.

In terms of the factors $\alpha_{XV\gamma}^{(i)}$ defined in
Eq.~(\ref{matrixvg}), one can express the partial width for the
radiative decay of $X(3872) \to \gamma V $ as
\begin{eqnarray}
\Gamma_{XV\gamma} & = & \frac{\alpha_{\rm em}}{10
m_{_X}^2}\Big[\sum_{i\leq
j=5}\alpha^{(i)}_{XV\gamma}\alpha^{(j)}_{XV\gamma}C^{\,ij}_{XV\gamma}\Big]P_{\gamma}^{\ast}
, \label{widthvg}
\end{eqnarray}
where $P_{\gamma}^{\ast}$ is the three-momentum of the decay
products. The coefficients $C^{\, ij}_{XV\gamma}$ are functions of
$m_{_X}$ and $m_{V}$. We give their explicit forms in
Appendix~\ref{app:cijrad}. To get this partial width, we have
applied the sum of the polarization vector for tensor
meson~\cite{LopezCastro:1997im}
\begin{eqnarray}
\sum_{\rm polar}
\epsilon_{\mu_1\nu_1}(p)\epsilon_{\mu_2\nu_2}^{\ast}(p) & = &
\frac{1}{2}\Big(\theta_{\mu_1\mu_2}\theta_{\nu_1\nu_2} +
\theta_{\mu_1\nu_2}\theta_{\nu_1\mu_2}\Big) -
\frac{1}{3}\theta_{\mu_1\nu_1}\theta_{\mu_2\nu_2} \; ,
\label{sumtensorvect}
\end{eqnarray}
where $\theta_{\mu\nu} = -g_{\mu\nu} + (p_\mu p_\nu)/m_{_X}^2$.

\begin{figure}[htbp]
\begin{center}
\includegraphics[scale=0.6]{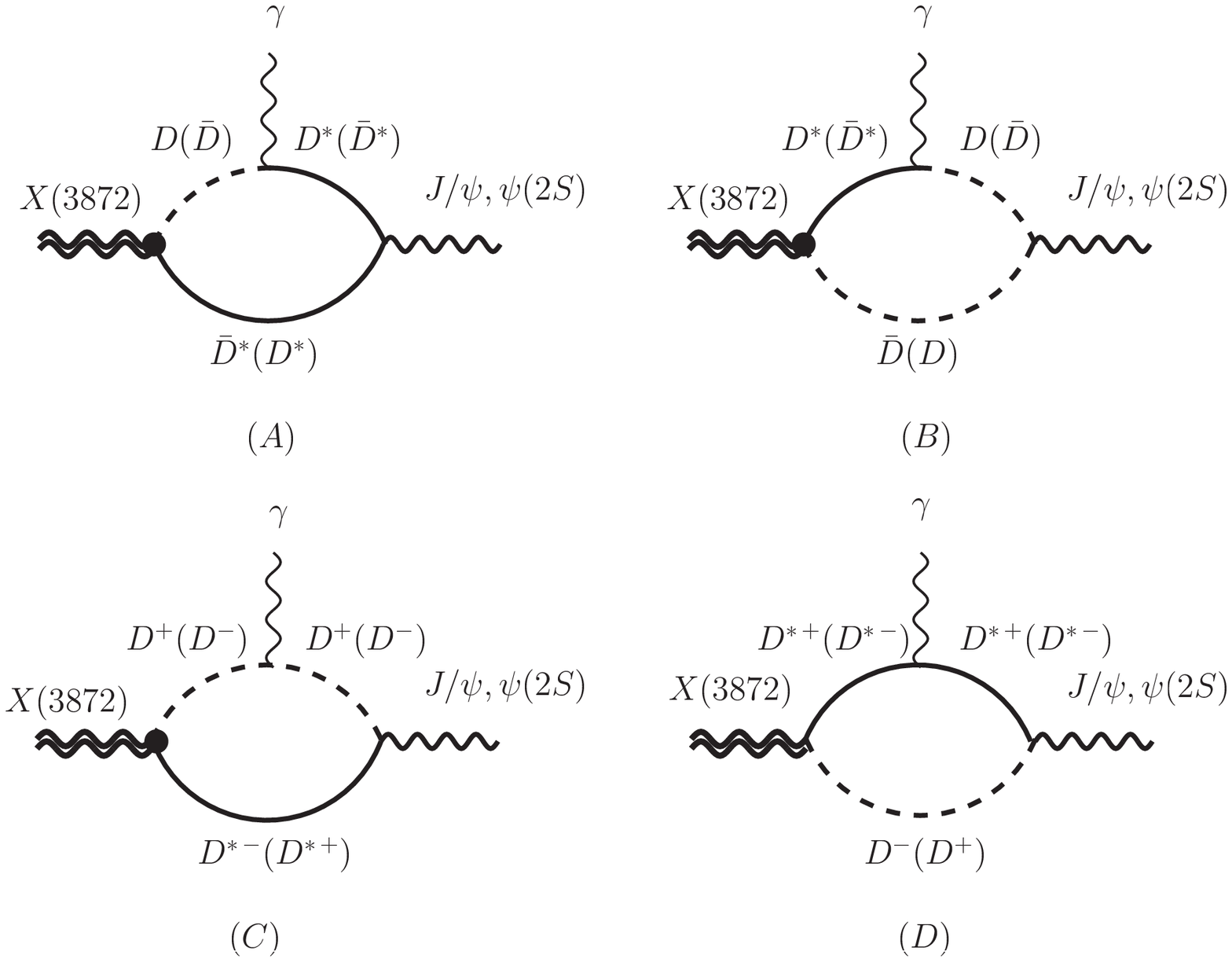}\\
\includegraphics[scale=0.6]{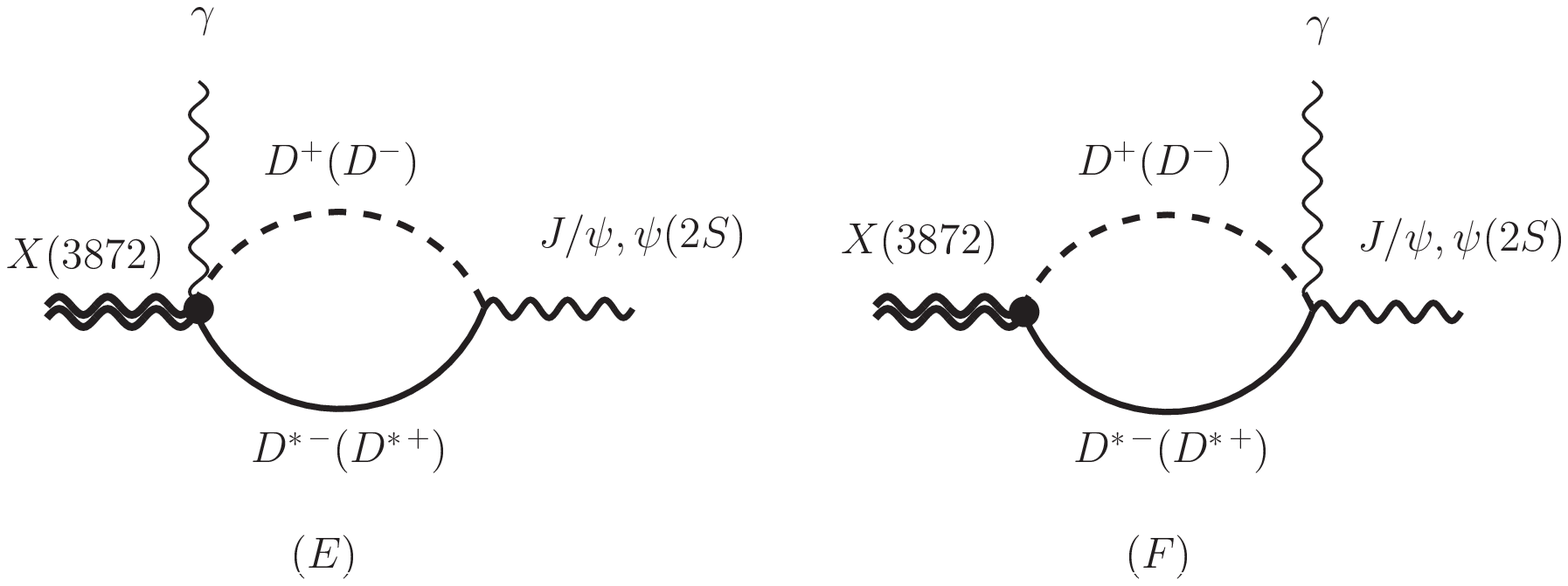}
\end{center}
\caption[Diagrams contributing to the radiative decays of $X(3872) \to \gamma J/\psi$, $\gamma\psi(2S)$.]{%
Diagrams contributing to the radiative decays of $X(3872) \to \gamma
J/\psi$, $\gamma\psi(2S)$. } \label{fig:vgamma}
\end{figure}

In our present work we study the radiative decays of $X \to \gamma
J/\psi$ and $X \to \gamma \psi(2S)$. The relevant diagrams are
illustrated in Fig.~\ref{fig:vgamma}.
After standard computation, we can yield the numerical results for
the partial widths. Both of them depend on the size parameter
$\Lambda_{_X}$ and the mixing angle $\phi$, or equivalently,
$\theta$.

\subsection{Strong $X(3872) \to J/\psi h$ decays}

Next, we study the strong decays of $X(3872) \to J/\psi h$ with $h$
as light pseudoscalar mesons $\pi^+\pi^-$ or $\pi^+\pi^-\pi^0$.
Measurements of the invariant masses of pions indicate that these
processes happen through $X(3872) \to J/\psi\rho$ for the $X(3872)
\to J/\psi\pi^+\pi^-$
decay~\cite{Choi:2003ue,Aubert:2004ns,Abulencia:2005zc} but $X(3872)
\to J/\psi\omega$ for the $X(3872) \to J/\psi\pi^+\pi^-\pi^0$
decay~\cite{Abe:2005ix,al.:2010jr}. So that in the following
calculation, we only take into account the decays via light vector
meson exchange. Explicitly, we consider the processes $X(3872) \to
J/\psi \rho^0 \to J/\psi \pi^+\pi^-$ and $X(3872) \to J/\psi\omega
\to J/\psi \pi^+\pi^-\pi^0$. Generally, one can write the matrix
elements of these decays as
\begin{eqnarray}
iM(X \to J/\psi h) & = & iM^\alpha(X \to J/\psi V)\frac{1}{q_1^2 -
m_V^2 + im_{_V}\Gamma_V}(g_{\alpha\sigma} - \frac{q_{1;\alpha}
q_{1;\sigma}}{m_{_V}^2})M^\sigma(V \to h), \label{generalmatrix}
\end{eqnarray}
where $V$ is the intermediate light vector meson with $m_{_V}$ and
$\Gamma_V$ as its mass and total width respectively. $q_1$ is the
momentum for the intermediate light vector meson $V$.

$M^\sigma(V \to h)$ stands for the contribution from the $V - h$
interaction which relates to the matrix element for $V \to h$ decay
via
\begin{eqnarray}
iM(V \to h) & = & i\epsilon_\sigma M^\sigma(V \to h),
\end{eqnarray}
with $\epsilon_\sigma$ as the polarization vector of the vector
meson.

\begin{figure}[htbp]
\begin{center}
\includegraphics[scale=0.6]{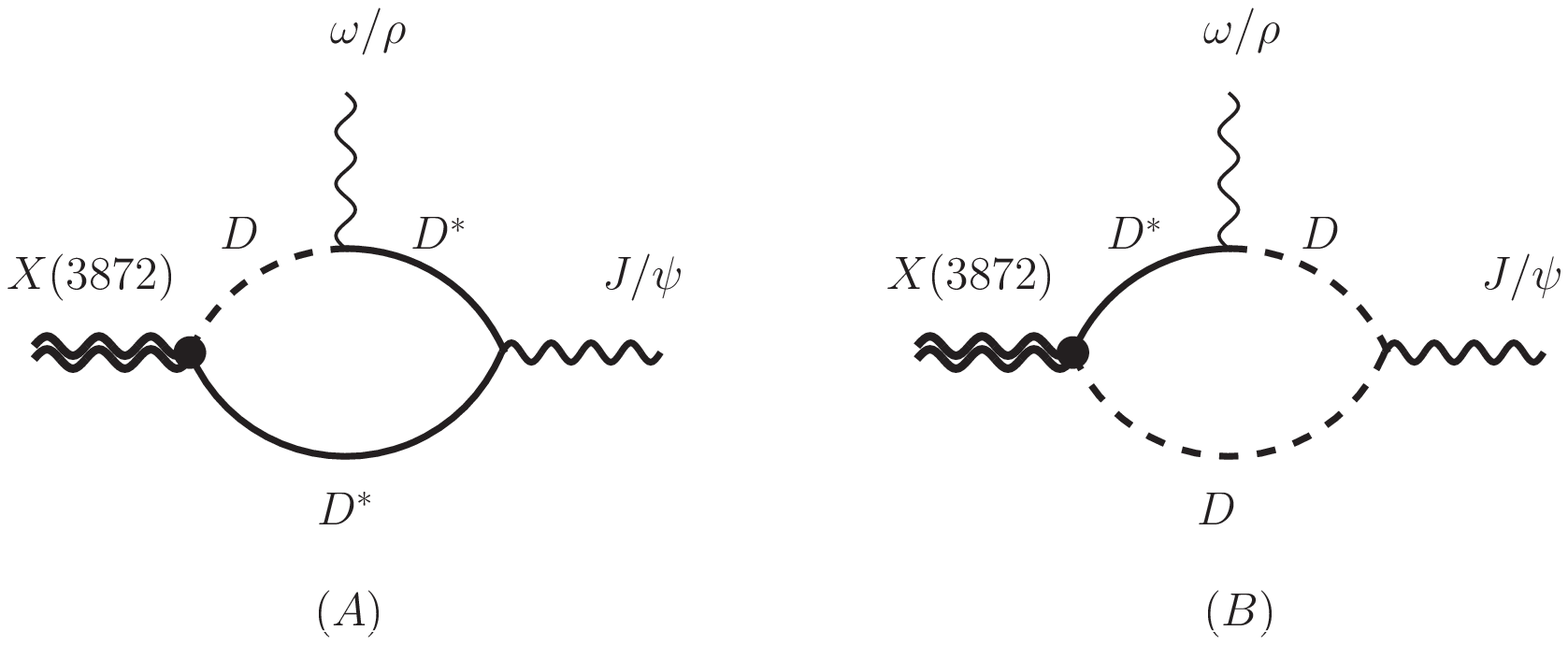}\\
\includegraphics[scale=0.6]{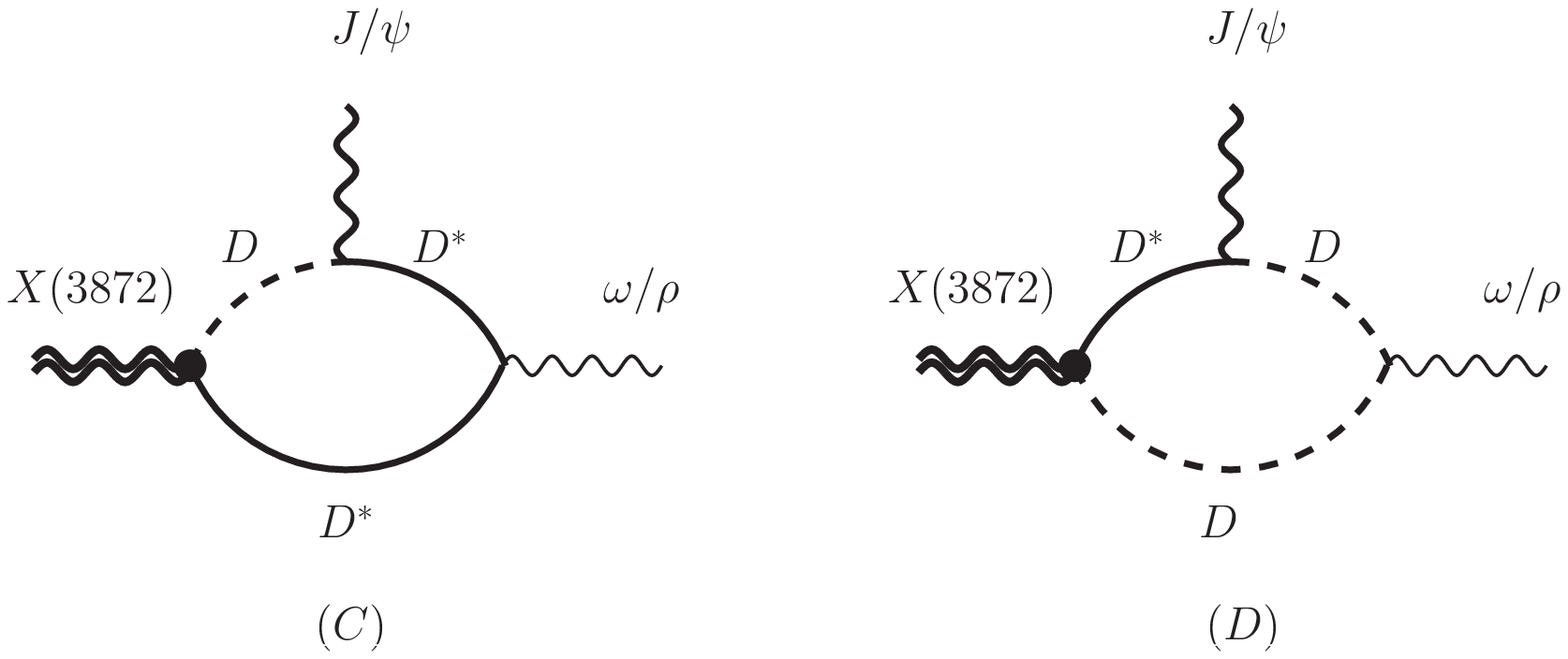}
\end{center}
\caption[Diagrams contributing to the strong decays of $X(3872) \to J/\psi\omega \to J/\psi \pi^+\pi^-\pi^0$, $J/\psi\rho \to J/\psi \pi^+\pi^-$.]{%
Diagrams contributing to the strong decays of $X(3872) \to
J/\psi\omega \to J/\psi \pi^+\pi^-\pi^0$, $J/\psi\rho \to J/\psi
\pi^+\pi^-$. } \label{fig:vv}
\end{figure}

The matrix element $M^\alpha(X \to J/\psi V)$ relates to the matrix
element for the decay $X \to J/\psi V$. Taking into account the
transverseness of the polarization vectors and symmetric property
and traceless of $\epsilon_{\mu\nu}$, we write the general form of
$M^\alpha(X \to J/\psi V)$ as
\begin{eqnarray}
iM^\alpha(X \to J/\psi  V) & = & iM_{X\psi
V}^{\mu\nu\alpha\beta}(m_{_X}^2,
m_{J/\psi}^2,q^2)\epsilon_{\mu\nu}(p)\epsilon_\beta^{\ast}(q_2) \nonumber\\
& = & i\Big\{G_{X\psi V}^{(1)}
\Big[\epsilon^{\mu\sigma\alpha\rho}g^{\nu\beta}p_\sigma q_\rho +
(\mu \leftrightarrow \nu) \Big] + G_{X\psi
V}^{(2)} \Big[\epsilon^{\mu\sigma\rho\beta}g^{\nu\alpha}p_\sigma q_\rho + (\mu \leftrightarrow \nu) \Big] \nonumber\\
& & + G_{X\psi V}^{(3)} \Big[\epsilon^{\mu\sigma\alpha\beta}p_\sigma
q^\nu + (\mu \leftrightarrow \nu) \Big] + G_{X\psi
V}^{(4)} \Big[\epsilon^{\mu\sigma\alpha\beta}q_\sigma q^\nu + (\mu \leftrightarrow \nu) \Big] \nonumber\\
& & + G_{X\psi V}^{(5)} \Big[\epsilon^{\mu\sigma\alpha\rho}p_\sigma
q_\rho q^\nu q^\beta + (\mu \leftrightarrow \nu) \Big] + G_{X\psi
V}^{(6)} \Big[\epsilon^{\mu\sigma\rho\beta}p_\sigma q_\rho q^\nu
q^\alpha + (\mu \leftrightarrow \nu) \Big] \nonumber\\
& & + G_{X\psi V}^{(7)}
\Big[\epsilon^{\sigma\rho\alpha\beta}p_\sigma q_\rho q^\mu q^\nu
\Big]\Big\} \epsilon_{\mu\nu}(p)\epsilon_{\beta}^{\ast}(q_2),
\label{matrixvv}
\end{eqnarray}
where in this expression, $p$ is the momentum of $X(3872)$ and $q =
q_1-q_2$ with $q_1$ and $q_2$ as the momentum of the intermediate
light vector meson $V$ and the final vector meson $J/\psi$
respectively. To write down this general form, we have considered
that both $M^\mu(\rho^0 \to \pi^+\pi^-)$ and $M^\mu(\omega \to
\pi^+\pi^-\pi^0)$ are transverse to $q_{1}^{\mu}$. In our
calculation of the form factors $G_{X\psi V}^{(i)}~ (i = 1,2,\ldots,
7)$ we have taken into account the diagrams illustrated in
Fig.~\ref{fig:vv}.

Using Eq.~(\ref{generalmatrix}) one can write the differential decay
rate into $J/\psi h$ as a function of the invariant mass $q_1$ of
the hadron $h$ as
\begin{eqnarray}
\frac{d \Gamma}{dq_1}(X \to J/\psi h) & = & \frac{1}{80\pi^2
m_{_X}^3}\frac{q_1 \lambda^{1/2}(m_{_X}^2,m_\psi^2,m_{_V}^2)}{(q_1^2
- m_{_V}^2)^2 + m_{_V}^2\Gamma_V^2}\Big(\sum_{i\leq
j=1}^7C_{ij}G_{X\psi V}^{(i)}G_{X\psi V}^{(j)}\Big)\Big(\int d\Pi
F(V \to h)\Big),
\end{eqnarray}
where the physical region of $q_1$ is $m_h \leq q_1 \leq m_X-m_\psi$
and the coefficients $C_{ij}$ are given in
Appendix~\ref{app:cijstr1}. $\int d\Pi F(V \to h)$ is the Lorentz
invariant phase space integral for $V \to h$ decay. For the $\rho^0
\to \pi^+ \pi^-$ decay we have
\begin{eqnarray}
\int d\Pi  F(\rho^0 \to \pi^+\pi^-) & = &
-\frac{g_{\rho\pi\pi}^2}{24\pi}\frac{1}{q_1}(q_1^2-4m_{\pi}^2)^{3/2},
\end{eqnarray}
with $g_{\rho\pi\pi}$ determined by the partial width for $\rho^{0}
\to \pi^+\pi^-$ decay as
\begin{eqnarray}
g_{\rho\pi\pi} = 6.0 \pm 0.03.\label{Gvpp}
\end{eqnarray}

In the case of $\omega \to \pi^+ \pi^- \pi^0$ decay, this process
has been studied in
Refs.~\cite{Fujiwara:1984mp,Harada:2003jx,Kaymakcalan:1983qq} using
the hidden local symmetry formulism. But in this formulism, the
matrix element depends on two unknown parameter combinations,
$c_1-c_2-c_3$ and $c_3$, explicitly,
\begin{eqnarray}
iM^\mu(\omega \to \pi^+ \pi^- \pi^0) & = & - g\frac{N_c}{16\pi^2
F_\pi^3}\epsilon^{\mu\nu\alpha\beta}q_{0;\nu}q_{+;\alpha}q_{-;\beta}\nonumber\\
& & \times
\Big\{3(c_1-c_2-c_3)+2c_3\Big[D_\rho((q_++q_-)^2)+D_\rho((q_-+q_0)^2)+D_\rho((q_0+q_+)^2)\Big]\Big\},
\end{eqnarray}
where $g = 5.80 \pm 0.91$ is the coupling constant introduced in the
gauge of the hidden local symmetry and $D_\rho(q^2)$ is the
propagator for $\rho$ meson which is normalized as $D_\rho(0) = 1$
and at the leading order
\begin{eqnarray}
D_\rho(q^2) & = & \frac{m_\rho^2}{m_\rho^2-q^2}.
\end{eqnarray}
Since it is a formidable work to determine these two combinations in
this paper we use the values determined in
Ref.~\cite{Braaten:2005ai} which are provided in
Appendix~\ref{app:omega3pi}.

\subsection{Strong $X(3872) \to \chi_{cJ}\pi^0$ decays}

\label{subsec:onepiondecay}

\begin{figure}[htbp]
\begin{center}
\includegraphics[scale=0.6]{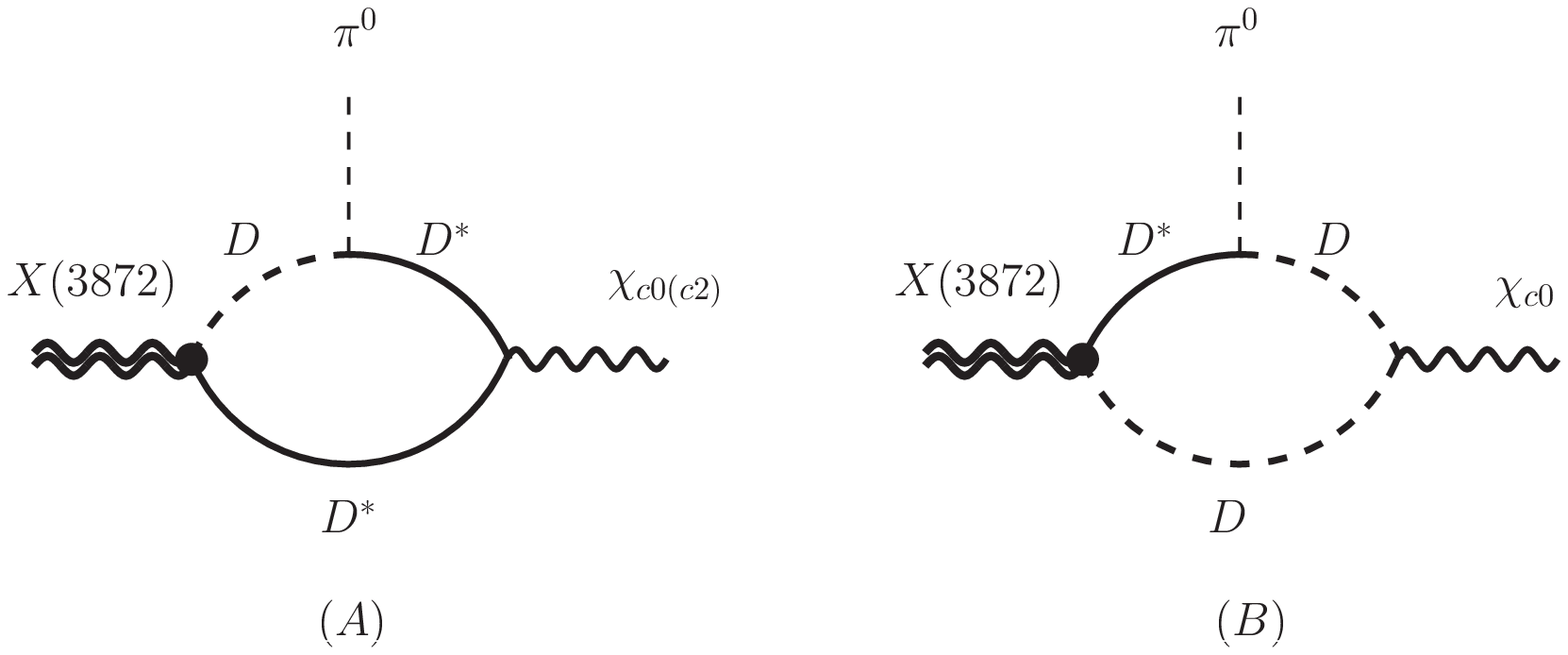}
\includegraphics[scale=0.6]{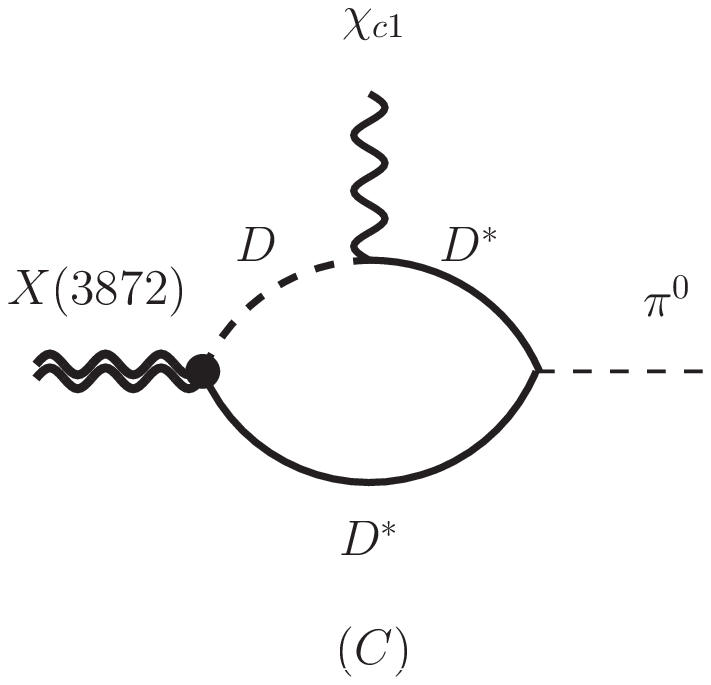}
\end{center}
\caption[Diagrams contributing to the strong decays of $X(3872) \to \chi_{cJ}\pi^0$.]{%
Diagrams contributing to the strong decays of $X(3872) \to
\chi_{cJ}\pi^0$. } \label{fig:chip}
\end{figure}

The decays $X(3872) \to \chi_{cJ}\pi^0~(J = 0,1,2)$ are important
and the measurement of these processes can discriminate between the
charmonium and the molecular interpretation of $X(3872)$ in the case
of $J^{PC} = 1^{++}$~\cite{Dubynskiy:2007tj}. We expect that this
holds for $J^{PC} = 2^{-+}$ case. In this subsection, we devote to
study these processes. Generally, the matrix elements for these
processes can be written as
\begin{eqnarray}
iM(X \to \chi_{c0}\pi^0) & = & iM_{X \chi_{c0}\pi^0}^{\mu\nu}
\epsilon_{\mu\nu}(p) = iG_{X \chi_{c0}\pi^0} q^\mu q^\nu
\epsilon_{\mu\nu}(p), \label{matrixps}\\
%%%%%%%%%%%%%%%%%%%%%%%%%%%%%%%%%%%%%%%%%%%%%%%%%%%%%%%%%%%%%%%%%%%%
iM(X \to \chi_{c1}\pi^0) & = &
iM_{X\chi_{c1}\pi^0}^{\mu\nu\alpha}\epsilon_{\mu\nu}(p)\epsilon_\alpha^{\ast}(q)
= iG_{X\chi_{c1}\pi^0}\Big[ \epsilon^{\mu\alpha\theta\tau}q^\nu +
\epsilon^{\nu\alpha\theta\tau}q^\mu \Big]
p_\theta q_\tau\epsilon_{\mu\nu}(p)\epsilon_\alpha^{\ast}(q) , \label{matrixpa}\\
%%%%%%%%%%%%%%%%%%%%%%%%%%%%%%%%%%%%%%%%%%%%%%%%%%%%%%%%%%%%%%%%%%%%
iM(X \to \chi_{c2}\pi^0) & = & iM_{X\chi_{c2}\pi^0}^{\mu\nu\alpha}\epsilon_{\mu\nu}(p)\epsilon_{\alpha\beta}^{\ast}(q) \nonumber\\
& = & i \Big\{G_{X\chi_{c2}\pi^0}^{(1)} q^\mu q^\nu p^\alpha p^\beta
+ G_{X\chi_{c2}\pi^0}^{(2)}(g^{\mu\alpha} g^{\nu\beta} +
g^{\nu\alpha}g^{\mu\beta}) \nonumber\\
& & + G_{X\chi_{c2}\pi^0}^{(3)}\Big[(g^{\mu\alpha}q^\nu +
g^{\nu\alpha}q^\mu)p^\beta + (\alpha \leftrightarrow
\beta)\Big]\Big\}\epsilon_{\mu\nu}(p)\epsilon_{\alpha\beta}^{\ast}(q)
,\label{matrixpt}
\end{eqnarray}
where in Eqs.~(\ref{matrixpa}) and (\ref{matrixpt}) $q$ is the
momentum of $\chi_{c1}$ and $\chi_{c2}$ but in Eq.~(\ref{matrixps})
$q=q_1-q_2$ with $q_1$ and $q_2$ as the momenta of the final states.
In our explicit calculation of the effective coupling constants
$G_{X \chi_{cJ}\pi^0}$, we include the diagrams illustrated in
Fig.~\ref{fig:chip}.

With the general expressions for the matrix elements, one can
express the general forms of the partial widths as
\begin{eqnarray}
\Gamma(X \to \chi_{c0}\pi^0) & = & \frac{G_{X \chi_{c0}\pi^0}^2}{60\pi m_{_X}^6}\lambda^{2}(m_{_X}^2,m_{\chi_{c0}}^2,m_{\pi^0}^2)P^{\ast}_{\chi_{c0}} , \label{widthps}\\
%%%%%%%%%%%%%%%%%%%%%%%%%%%%%%%%%%%%%%%%%%%%%%%%%%%%%%%%%%%%%%%%%%%%
\Gamma(X \to \chi_{c1}\pi^0) & = & \frac{G_{X \chi_{c1}\pi^0}^2}{160\pi m_{_X}^4}\lambda^{2}(m_{_X}^2,m_{\chi_{c1}}^2,m_{\pi^0})P^{\ast}_{\chi_{c1}} , \label{widthpa}\\
%%%%%%%%%%%%%%%%%%%%%%%%%%%%%%%%%%%%%%%%%%%%%%%%%%%%%%%%%%%%%%%%%%%%
\Gamma(X \to \chi_{c2}\pi^0) & = & \frac{1}{40\pi
m_{_X}^2}\Big[\sum_{i\leq j=1}^3G_{X \chi_{c2}\pi^0}^{(i)}G_{X
\chi_{c2}\pi^0}^{(j)}C^{ij}_{X\chi_{c2}\pi^0}\Big]P^{\ast}_{\chi_{c2}}
,\label{widthptt}
\end{eqnarray}
where $\lambda(a^2,b^2,c^2) =
\big(a^2-(b-c)^2\big)\big(a^2-(b+c)^2\big)$ is the K$\ddot{a}$llen
function and $P^{\ast}_{\chi_{cJ}}$ is the three momentum of the
decay products. The coefficients $C^{ij}_{X\chi_{c2}\pi^0}$ is given
in Appendix \ref{app:cijstr2}.

\section{Numerical results and discussions}

\label{sec:numerical}

\subsection{Parameter fitting}

\label{subsec:numericalparameter}

\begin{figure}[htbp]
\begin{center}
\includegraphics[scale=0.95]{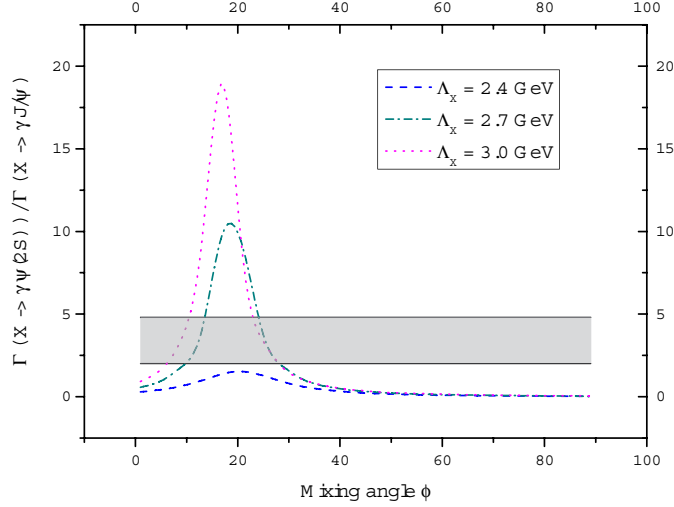}
\end{center}
\caption[Mixing angle $\phi$ dependence of the branching ratio $\Gamma(X \to \gamma\psi(2S))/\Gamma(X \to \gamma J/\psi)$ in the case of $\Delta E = 0.5$~MeV. The grey band corresponds to the data with the error bar.]{%
Mixing angle $\phi$ dependence of the branching ratio $\Gamma(X \to
\gamma\psi(2S))/\Gamma(X \to \gamma J/\psi)$ in the case of $\Delta
E = 0.5$~MeV. The grey band corresponds to the data with the error
bar.} \label{fig:emratio}
\end{figure}

In the present analysis, because of the uncertainty of the $X(3872)$
mass measurement, we take the binding energy defined in
Eq.~(\ref{eq:bindenergy}) as $\Delta E = 0.5, 1.0$ and $1.5~$MeV.
Then in our model, there are two parameters, i.e., the mixing angle
$\phi$ between the isospin singlet and isospin triplet components
and the size parameter $\Lambda_{_X}$. In the numerical calculation,
since we have no prior information from QCD, we take these two
parameters as free ones to fit the following data
\begin{eqnarray}
\frac{\mathcal{B}(X \to \gamma \psi(2S))}{\mathcal{B}(X \to \gamma
J/\psi)} & = & 3.4 \pm 1.4 ~\mbox{\cite{:2008rn}}, \label{dataratioem}\\
\frac{\mathcal{B}(X(3872) \to
J/\psi\pi^+\pi^-\pi^0)}{\mathcal{B}(X(3872) \to J/\psi\pi^+\pi^-)} &
= & 1.0 \pm 0.4({\rm stat.}) \pm 0.3 ({\rm syst.})
~\mbox{\cite{Abe:2005ix}}, \label{eq:data3pto2p}\\
\frac{\mathcal{B}(X(3872) \to \gamma J/\psi)}{\mathcal{B}(X(3872)
\to J/\psi\pi^+\pi^-)} & = & 0.14 \pm 0.05~~~~~~~~{\rm
Belle}~\mbox{\cite{Abe:2005ix}},\nonumber\\
& & 0.33 \pm 0.12 ~~~~~~~~{\rm
BaBar~\mbox{\cite{:2008rn}}}.\label{gatagamma2pi}
\end{eqnarray}
We scan the mixing angle $\phi$ from $0^\circ$ to $90^\circ$ with a
step of $1^\circ$ and the size parameter $\Lambda_{_X}$ from $0.1$
to $3.0~$GeV with a step of $0.1~$GeV. As an illustration, we plot
the mixing angle $\phi$ dependence of the branching ratio $\Gamma(X
\to \gamma\psi(2S))/\Gamma(X \to \gamma J/\psi)$ at three typical
values of $\Lambda_{_X} = 2.4, 2.7$ and $3.0$~GeV in the case of
$\Delta E = 0.5$~MeV in Fig.~\ref{fig:emratio}. This figure shows
that any values for $\phi$ cannot yield the ratio for $\Lambda_{_X}
= 2.4~$GeV case, while the range for $\phi$ is constrained for
$\Lambda_{_X} = 2.7,3.0$~GeV. In this way we determine the range of
the two parameters using the data in
Eqs.~(\ref{dataratioem},\ref{eq:data3pto2p},\ref{gatagamma2pi}).

In Table.~\ref{table:fit}, our results of these two parameters which
can reproduce the data within $1\sigma$ together with the values of
relevant branching ratios are given. As can be seen, we can
reproduce the data in the first two columns and the Babar data in
the third column quite well but we cannot get the result of Belle.
Also listed in Table.~\ref{table:fit} are the values of the ratio
$\mathcal{B}(X(3872) \to \gamma \psi(2S))/\mathcal{B}(X(3872) \to
J/\psi \pi^+\pi^-)$ which was measured as
\begin{eqnarray}
\frac{\mathcal{B}(X(3872) \to \gamma \psi(2S))}{\mathcal{B}(X(3872)
\to J/\psi \pi^+\pi^-)} & = & 1.1 \pm 0.4, ~~~~~~~~{\rm
BaBar~\mbox{\cite{:2008rn}}}.
\end{eqnarray}
%
%
%For a certain value of $\Lambda_{_X}$, the mixing angle $\phi$
%dependence of the branching ratio can be understood from the values
%of mixing angle $\theta$. For simplicity consider the isospin
%symmetric case.
\begin{table}

\caption{\label{table:fit}  Fitted parameters $\theta$ and
$\Lambda_{_X}$ and the corresponding branching ratio.}

\begin{tabular}{lllllll}
\hline \hline \hspace*{.1cm} $\phi$ \hspace*{.2cm} & \hspace*{.1cm}
$\Lambda_{_X}$(GeV) \hspace*{.1cm} & \hspace*{.1cm} $\Delta E$(MeV)
\hspace*{.1cm}& \hspace*{.3cm} $\frac{\mathcal{B}(X \to \gamma
\psi(2S))}{\mathcal{B}(X \to \gamma J/\psi)}$\hspace*{.1cm} &
\hspace*{.2cm} $\frac{\mathcal{B}(X \to
J/\psi\pi^+\pi^-\pi^0)}{\mathcal{B}(X \to
J/\psi\pi^+\pi^-)}$\hspace*{.1cm} & \hspace*{.2cm}
$\frac{\mathcal{B}(X \to \gamma J/\psi)}{\mathcal{B}(X \to
J/\psi\pi^+\pi^-)}$ \hspace*{.2cm}  & \hspace*{.2cm}
$\frac{\mathcal{B}(X \to \gamma\psi(2S))}{\mathcal{B}(X \to
J/\psi\pi^+\pi^-)}$ \hspace*{.2cm} \\
\hline
\,\, $12^\circ $ & \,\,\, $ 2.6 $ \,\,\,& \,\,\, $ 0.5 $ \,\,\,& \,\,\,\,\,\,\,\, $2.089$ & \,\,\,\, $0.483$\,\,\,& \,\,\,\, $0.218$\,\,\, & \,\,\,\, $0.455$\,\,\, \\
%\hline
\,\, ~& \,\,\, ~\,\,\,& \,\,\, $ 1.0 $ \,\,\,& \,\,\,\,\,\,\,\, $2.101$ & \,\,\,\, $0.461$\,\,\,& \,\,\,\, $0.214$\,\,\, & \,\,\,\, $0.450$\,\,\, \\
%\hline
\,\, ~ & \,\,\, ~ \,\,\,& \,\,\, $ 1.5 $ \,\,\,& \,\,\,\,\,\,\,\, $2.113$ & \,\,\,\, $0.439$\,\,\,& \,\,\,\, $0.210$\,\,\, & \,\,\,\, $0.444$\,\,\, \\
\hline
\,\, $11^\circ $ & \,\,\, $ 2.7 $ \,\,\, &\,\,\, $ 0.5 $ \,\,\,& \,\,\,\,\,\,\,\, $2.628$ & \,\,\,\, $0.591$\,\,\,& \,\,\,\, $0.211$\,\,\, & \,\,\,\, $0.555$\,\,\, \\
%\hline
\,\, ~ & \,\,\, ~ \,\,\, &\,\,\, $ 1.0 $ \,\,\,& \,\,\,\,\,\,\,\, $2.640$ & \,\,\,\, $0.566$\,\,\,& \,\,\,\, $0.208$\,\,\, & \,\,\,\, $0.549$\,\,\, \\
%\hline
\,\, ~ & \,\,\, ~ \,\,\, &\,\,\, $ 1.5 $ \,\,\,& \,\,\,\,\,\,\,\, $2.647$ & \,\,\,\, $0.542$\,\,\,& \,\,\,\, $0.206$\,\,\, & \,\,\,\, $0.545$\,\,\, \\
\hline
\,\, $10^\circ $ & \,\,\, $ 2.7-2.8 $ \,\,\, &\,\,\, $ 0.5 $ \,\,\,& \,\,\,\,\,\,\,\, $2.158 - 3.007$ & \,\,\,\, $0.748 - 0.740$\,\,\,& \,\,\,\, $0.320 - 0.233$\,\,\, & \,\,\,\, $0.690 - 0.700$\,\,\, \\
%\hline
\,\, ~ & \,\,\, ~ \,\,\, &\,\,\, $ 1.0 $ \,\,\,& \,\,\,\,\,\,\,\, $2.165 - 3.014$ & \,\,\,\, $0.716 - 0.705$\,\,\,& \,\,\,\, $0.316 - 0.229$\,\,\, & \,\,\,\, $0.684 - 0.690$\,\,\, \\
%\hline
\,\, ~ & \,\,\, ~ \,\,\, &\,\,\, $ 1.5 $ \,\,\,& \,\,\,\,\,\,\,\, $2.168 - 3.017$ & \,\,\,\, $0.685 - 0.673$\,\,\,& \,\,\,\, $0.312 - 0.226$\,\,\, & \,\,\,\, $0.676 - 0.682$\,\,\, \\
\hline
\,\, ~ & \,\,\,  \,\,\, &\,\,\, $ 0.5$ \,\,\,& \,\,\,\,\,\,\,\, $2.430 - 3.486$ & \,\,\,\, $0.964 - 1.035$ \,\,\,& \,\,\,\, $0.369 - 0.283$ \,\,\, & \,\,\,\, $0.897 - 0.987$ \,\,\, \\
%\hline
\,\, $9^\circ$ & \,\,\, $ 2.8 - 3.0$ \,\,\, &\,\,\, $ 1.0 $ \,\,\,& \,\,\,\,\,\,\,\, $2.433 - 3.504$ & \,\,\,\, $0.917 - 0.971$ \,\,\,& \,\,\,\, $0.363 - 0.274$ \,\,\, & \,\,\,\, $0.883 - 0.960$ \,\,\, \\
%\hline
\,\, ~ & \,\,\, ~ \,\,\, &\,\,\, $ 1.5$ \,\,\,& \,\,\,\,\,\,\,\, $2.433 - 3.516$ & \,\,\,\, $0.874 - 0.914$ \,\,\,& \,\,\,\, $0.358 - 0.266$ \,\,\, & \,\,\,\, $0.871 - 0.935$ \,\,\, \\
\hline
\,\, data & \,\,\, ~ \,\,\, &\,\,\, ~ \,\,\,& \,\,\,\,\,\,\,\, $3.4 \pm 1.4$~\cite{:2008rn} & \,\,\,\, $1.0 \pm 0.4 \pm 0.3 $~\cite{Abe:2005ix} \,\,\,& \,\,\,\, $0.14 \pm 0.05 $~\cite{Abe:2005ix} \,\,\, & \,\,\,\, $1.1 \pm 0.4$~\cite{:2008rn} \,\,\, \\
%\hline
\,\, ~ & \,\,\, ~ \,\,\, &\,\,\, ~ \,\,\,& \,\,\,\,\,\,\,\, ~ & \,\,\,\, ~ \,\,\,& \,\,\,\, $0.33 \pm 0.12 $~\cite{:2008rn} \,\,\, & \,\,\,\, ~ \,\,\, \\
\hline \hline
\end{tabular}
%\end{center}
\end{table}
Taking into account this ratio, one can see that the smaller values
of the mixing angle $\phi$ are preferred. So in the following, we
restrict ourselves to the case with $\phi = 9^\circ$.

In Table.~\ref{table:coupling} we present our numerical results of
the relevant coupling constants $g_{_X}^{N}$ and $g_{_X}^{C}$. From
these results we conclude that the effective coupling constants are
stable against the size parameter $\Lambda_{_X}$.
\begin{table}

\caption{\label{table:coupling}  Effective coupling constants
$g_{_X}^{N}$ and $g_{_X}^{C}$ from the fitted parameters $\phi$ and
$\Lambda_{_X}$.}

\begin{tabular}{lllllll}
\hline \hline \,\, $\phi$ \,\, & \,\, $\Lambda_{_X}$(GeV) \,\, &
\,\, $\Delta E$(MeV) \,\, & \,\,\,\,\,\,\,\,\,\,\,\,\,\,
$g_{_X}^{N}$ \,\, & \,\,\,\,\,\,\,\,\,\,\,\,\,\,
$g_{_{X}}^{C}$ \,\, \\
\hline
%\,\, $12^\circ $ & \,\,\, $ 2.6 $ \,\,\,& \,\,\, $ 0.5 $ \,\,\,& \,\,\,\,\,\,\,\, $17.20$ & \,\,\,\, $11.99$\,\,\,\\
%\hline
%\,\, ~& \,\,\, ~\,\,\,& \,\,\, $ 1.0 $ \,\,\,& \,\,\,\,\,\,\,\, $17.29$ & \,\,\,\, $12.03$\,\,\, \\
%\hline
%\,\, ~ & \,\,\, ~ \,\,\,& \,\,\, $ 1.5 $ \,\,\,& \,\,\,\,\,\,\,\, $17.39$ & \,\,\,\, $12.07$\,\,\, \\
%\hline
%\,\, $11^\circ $ & \,\,\, $ 2.7 $ \,\,\, &\,\,\, $ 0.5 $ \,\,\,& \,\,\,\,\,\,\,\, $16.84$ & \,\,\,\, $12.16$\,\,\, \\
%\hline
%\,\, ~ & \,\,\, ~ \,\,\, &\,\,\, $ 1.0 $ \,\,\,& \,\,\,\,\,\,\,\, $16.93$ & \,\,\,\, $12.20$\,\,\, \\
%\hline
%\,\, ~ & \,\,\, ~ \,\,\, &\,\,\, $ 1.5 $ \,\,\,& \,\,\,\,\,\,\,\, $17.02$ & \,\,\,\, $12.24$\,\,\, \\
%\hline
%\,\, $10^\circ $ & \,\,\, $ 2.7-2.8 $ \,\,\, &\,\,\, $ 0.5 $ \,\,\,& \,\,\,\,\,\,\,\, $16.64 - 16.50$ & \,\,\,\, $ 12.48 - 12.34 $\,\,\, \\
%\hline
%\,\, ~ & \,\,\, ~ \,\,\, &\,\,\, $ 1.0 $ \,\,\,& \,\,\,\,\,\,\,\, $16.73 - 16.58 $ & \,\,\,\, $12.52 - 12.38$\,\,\, \\
%\hline
%\,\, ~ & \,\,\, ~ \,\,\, &\,\,\, $ 1.5 $ \,\,\,& \,\,\,\,\,\,\,\, $16.81 - 16.66$ & \,\,\,\, $12.56 - 12.42$\,\,\, \\
%\hline
\,\, ~ & \,\,\, ~ \,\,\, &\,\,\,\,\,\,\,\,\,\,\, $ 0.5$ \,\,\,& \, $16.29 - 15.84$ & \, $12.64 - 12.40$ \,\,\, \\
%\hline
\,\, $9^\circ$ & \,\,\, $ 2.8 - 3.0$ \,\,\, &\,\,\,\,\,\,\,\,\,\,\, $ 1.0 $ \,\,\,& \, $16.38 - 15.94$ & \, $12.68 - 12.44$ \,\,\, \\
%\hline
\,\, ~ & \,\,\, ~ \,\,\, &\,\,\,\,\,\,\,\,\,\,\, $ 1.5$ \,\,\,& \, $16.46 - 16.04$ & \, $12.72 - 12.48$ \,\,\, \\
\hline \hline
\end{tabular}
%\end{center}
\end{table}
Once comparing with Ref.~\cite{Dong:2009yp}, one may find that our
results for the coupling constants are bigger than the corresponding
ones given there. This can be interpreted as the following: In our
case, $X(3872)$ has quantum numbers $J^{PC} = 2^{-+}$ so the
coupling between $X(3872)$ and its constituents $DD^{\ast}$ is in
$P-$wave. Compared with the case that $X(3872)$ has quantum numbers
$J^{PC} = 1^{++}$, it needs stronger attractive interaction to
compensate the repulsive interaction induced by angular momentum.

Our result of the small mixing angle implies that the isospin
singlet component is dominant. Explicitly, one can write the wave
function of $X(3872)$ in terms of the isospin basis as
\begin{eqnarray}
|X(3872)\rangle = 0.988 \times |X(3872)\rangle_{I=0} + 0.156 \times
|X(3872)\rangle_{I=1}. \label{eq:wavex}
\end{eqnarray}
From this one might think that decay $X \to J/\psi \pi^+\pi^-$ would
be strongly suppressed compared to $X \to J/\psi \pi^+\pi^-\pi^0$
decay. However, this is not the case. One can understand this as the
following: Concerning the fact that the mass of $X(3872)$ is around
the threshold of $J/\psi V$ and following Ref.~\cite{Braaten:2005ai}
one can simplify the matrix element (\ref{matrixvv}) into the
approximate form
\begin{eqnarray}
iM^\alpha(X \to J/\psi  V) & = & iG_{X\psi
V}\epsilon^{\mu\sigma\alpha\beta}p_\sigma q^\nu
\epsilon_{\mu\nu}(p)\epsilon_{\beta}^{\ast}(q_2).
\end{eqnarray}
So that, in terms of the effective coupling constants
$G_{X\psi\omega}$ and $G_{X\psi\rho}$ the ratio $\mathcal{B}(X \to
J/\psi \pi^+\pi^-\pi^0)/\mathcal{B}(X \to J/\psi \pi^+\pi^-)$ can be
expressed as
\begin{eqnarray}
\frac{\mathcal{B}(X \to J/\psi \pi^+\pi^-\pi^0)}{\mathcal{B}(X \to
J/\psi \pi^+\pi^-)} = % 0.0127
\frac{|G_{X\psi\omega}|^2}{|G_{X\psi\rho}|^2}\frac{\int d\Pi(X \to
J/\psi \pi^+\pi^-\pi^0)}{\int d\Pi(X \to J/\psi \pi^+\pi^-)},
\label{eq:ratiophase}
\end{eqnarray}
where $\int d\Pi$ denotes the phase space integral. If one naively
use the wave function (\ref{eq:wavex}) one can get
$|G_{X\psi\omega}|^2/|G_{X\psi\rho}|^2 \simeq (0.988/0.156)^2 \simeq
40 $. But this ratio is strongly suppressed in case one includes the
phase space factor, i.e., $\int d\Pi(X \to J/\psi
\pi^+\pi^-\pi^0)/\int d\Pi(X \to J/\psi \pi^+\pi^-) \simeq 0.0127$
which leads to $\mathcal{B}(X \to J/\psi
\pi^+\pi^-\pi^0)/\mathcal{B}(X \to J/\psi \pi^+\pi^-) \simeq 0.5$.
More precisely, without including the difference arising from the
loop integrals in Fig.~\ref{fig:vv}, one can yield
\begin{eqnarray}
\frac{G_{X\psi\omega}}{G_{X\psi\rho}} & \simeq & \frac{g_{_X}^{N} +
g_{_X}^{C}}{g_{_X}^{N}-g_{_X}^{C}} \cdot
\frac{g_{DD\omega}}{g_{DD\rho}} = 8.9 ,\label{eq:ratiosv}
\end{eqnarray}
where we have used the numerical results of
Table.~\ref{table:coupling} and Eq.~(\ref{eq:couplingddv}). From
this we yield $\mathcal{B}(X \to J/\psi
\pi^+\pi^-\pi^0)/\mathcal{B}(X \to J/\psi \pi^+\pi^-) \simeq 1.0$
which is consistent with the experimental data given in
(\ref{eq:data3pto2p}).

%Using $\phi = 9^\circ$, one can yield the magnitude of the neutral
%component and charged component mixing angle $\theta$ as $\theta =
%36^\circ$. So that the wave function of $X(3872)$ in terms of the
%neutral and charged basis can be expressed as
%\begin{eqnarray}
%|X(3872)\rangle =0.572 \times | D^{0}\bar{D}^{\ast \, 0}\rangle +
%0.416 \times | D^{+}D^{\ast \, -}\rangle + {\rm C.c.}.
%\end{eqnarray}

\subsection{Partial widths for decays with $J/\psi$ or $\psi(2S)$ in the final states}

In Table.~\ref{table:widthdata} we give partial widths for decays
with $J/\psi$ or $\psi(2S)$ in the final states. From this we see
the partial widths are of order of KeV.
\begin{table}

\caption{\label{table:widthdata}  Fitted parameters $\phi$ and
$\Lambda_{_X}$ and the corresponding branching ratio related to the
data.}

\begin{tabular}{lllllll}
\hline \hline \;\; $\phi$ \;\; & \;\;\;\;\;\; $\Lambda_{_X}$ \;\; &
\;\; $\Delta E$ \;\; & \;\; $\Gamma(X \to \gamma J/\psi)$ \;\; &
\;\; $\Gamma(X \to \gamma \psi(2S))$ \;\; & \;\; $\Gamma(X \to
J/\psi\pi^+\pi^-\pi^0)$ \;\; &
\;\; $\Gamma(X \to J/\psi\pi^+\pi^-)$ \;\; \\
\;\; ~ \;\; & \;\;\; (GeV) \;\; & \; (MeV) \;\; & \;\;\;\;\;\; (KeV)
\;\; & \;\;\;\;\;\; (KeV) \;\; & \;\;\;\;\;\;\;\;\;\; (KeV) \;\; &
\;\;\;\;\;\;\;\; (KeV) \;\; \\
\hline
\,\, ~ & \,\,\, ~ \,\,\, &\,\,\, $ 0.5$ \,\,\, & \,\,\,\, $2.085 - 1.872$ & \,\,\,\,\,\, $5.066 - 6.525$ \,\,\,& \,\,\,\,\,\,\,\,\,\,\,\,\,\, $5.447 - 6.842$ \,\,\, & \,\,\,\,\,\,\,\,\,\, $5.648 - 6.612$ \,\,\, \\
%\hline
\,\, $9^\circ$ & \,\,\, $ 2.8 - 3.0$ \,\,\, &\,\,\, $ 1.0$ \,\,\, & \,\,\,\, $2.082 - 1.864$ & \,\,\,\,\,\, $5.065 - 6.533$ \,\,\,& \,\,\,\,\,\,\,\,\,\,\,\,\,\, $5.253 - 6.606$ \,\,\, & \,\,\,\,\,\,\,\,\,\, $5.730 - 6.802$ \,\,\, \\
%\hline
\,\, ~ & \,\,\, ~ \,\,\, &\,\,\, $ 1.5$ \,\,\, & \,\,\,\, $2.079 - 1.859$ & \,\,\,\,\,\, $5.058 - 6.537$ \,\,\,& \,\,\,\,\,\,\,\,\,\,\,\,\,\, $5.067 - 6.380$ \,\,\, & \,\,\,\,\,\,\,\,\,\, $5.801 - 6.977$ \,\,\, \\
\hline \hline
\end{tabular}
%\end{center}
\end{table}
These partial widths have been computed in the case that $X(3872)$
has positive
parity~\cite{Braaten:2005ai,Dong:2009uf,Dong:2008gb,Dong:2009yp}.
For the strong decays $X(3872) \to J/\psi\pi^+\pi^-$ and $X(3872)
\to J/\psi\pi^+\pi^-\pi^0$, it was found that the decay widths are
around $50~$KeV which are both about one order larger than our
present results. But for radiative decays, the results depend on
model closely. In the case that $X(3872)$ with quantum numbers
$J^{PC} = 1^{++}$ is a bound state of mesons~\cite{Dong:2008gb}, the
radiative decay widths are of order KeV which are at the same order
as our present results, while if $X(3872)$ is a mixing state of
molecule (without the charged $DD^{\ast}$ components) and charmonium
component~\cite{Dong:2009yp}, the decay width of $X(3872) \to \gamma
J/\psi$ is found to be of $100-200~$KeV which is much larger than
that computed from the hadronic molecule assumption. In
Ref.~\cite{Dong:2009uf}, an admixture model of molecule with the
charged $DD^{\ast}$ components and the charmonium components was
applied to study the radiative decays, the similar results as ours
were yield.

The inclusion of other components as was done in the case that
$X(3872)$ has quantum numbers $J^{PC} = 1^{++}$ may change our
results. One possibility is to include $J/\psi\omega$ and
$J/\psi\rho$ in the wave function~\cite{Swanson:2003tb,Dong:2009yp}.
This may increase the magnitudes of the strong decay widths and the
results depend on the probability of $J/\psi\omega$ and $J/\psi\rho$
in $X(3872)$ even in the case that only the long distance effect is
considered. In this sense, if the strong decay widths for tensor
$X(3872)$ are observed bigger than our present results, one may
conclude that the tensor $X(3872)$ cannot be a pure $DD^{\ast}$
molecule and other component should be included. Another possibility
is to regard $X(3872)$ as a mixing state of $c\bar{c}$ and
$DD^{\ast}$. One may borrow the lesson from the $X(3872)$ with
$1^{++}$ case~\cite{Dong:2008gb} to naively expect that this change
of the wave function of $X(3872)$ may improve the magnitude of
radiative decay width of $X(3872)$. Of course, in case other
constituent is included, the probabilities of the relevant
components should be fitted from data again so the magnitudes of the
partial widths should be studied in detail.

From these discussions, we would like to stress that the precise
measurement of the strong decay widths can provide some clues on the
structure of $X(3872)$.

\subsection{Predictions of the $X(3872) \to \chi_{cJ}\pi$ decays}

As was discussed in subsection~\ref{subsec:onepiondecay}, we expect
that the measurement of the decays $X(3872) \to
\chi_{cJ}\pi^0~(J=0,1,2)$ can discriminate between the charmonium
and the molecular interpretation of $X(3872)$. With the expressions
given in subsection~\ref{subsec:onepiondecay} and the parameters
fixed in subsection~\ref{subsec:numericalparameter}, the decay
widths of $X(3872) \to \chi_{cJ}\pi^0$ can be calculated explicitly.
Table.~\ref{table:width1pi} is the summary of our numerical results.

\begin{table}

\caption{\label{table:width1pi}  Partial widths for $X(3872) \to
\chi_{cJ} \pi^0$ decays.}

\begin{tabular}{llllll}
\hline \hline \;\; $\phi$ \;\; & \;\;\;\; $\Lambda_{_X}$ \;\; & \;\;
$\Delta E$ \;\; & \;\; $\Gamma(X \to \chi_{c0} \pi^0)$ \;\; &
\hspace*{.2cm} $\Gamma(X \to \chi_{c1} \pi^0)$\hspace*{.1cm} &
\hspace*{.2cm}
$\Gamma(X \to \chi_{c2} \pi^0)$ \hspace*{.2cm}  \\
%\hline
\hspace*{.1cm} ~ \hspace*{.2cm} & \;\; (GeV) \;\; & \; (MeV) \;\; &
\;\;\;\;\;\; (KeV) \;\; & \;\;\;\;\;\; (KeV) \;\; & \;\;\;\;\;\;
(KeV) \;\;  \\
\hline
%\,\, $12^\circ $ & \,\,\, $ 2.6 $ \,\,\, & \,\,\, $ 0.5 $ \,\,\, & \,\,\,\,\,\,\,\, $37.40~$KeV & \,\,\,\, $0.499~$KeV \,\,\,& \,\,\,\, $901.6~$KeV \,\,\,\\
%\hline
%\,\, ~ & \,\,\, ~ \,\,\, & \,\,\, $ 1.0 $ \,\,\, & \,\,\,\,\,\,\,\, $37.71~$KeV & \,\,\,\, $0.502~$KeV \,\,\,& \,\,\,\, $919.2~$KeV \,\,\,\\
%\hline
%\,\, ~ & \,\,\, ~ \,\,\, & \,\,\, $ 1.5 $ \,\,\, & \,\,\,\,\,\,\,\, $38.02~$KeV & \,\,\,\, $0.505~$KeV \,\,\,& \,\,\,\, $936.2~$KeV \,\,\,\\
%\hline
%\,\, $11^\circ $ & \,\,\, $ 2.7 $ \,\,\,& \,\,\, $ 0.5 $ \,\,\,& \,\,\,\,\,\,\,\, $32.48~$KeV & \,\,\,\, $0.4311~$KeV\,\,\,& \,\,\,\, $546.8~$KeV \,\,\,\\
%\hline
%\,\, ~ & \,\,\, ~ \,\,\,& \,\,\, $ 1.0 $ \,\,\,& \,\,\,\,\,\,\,\, $32.76~$KeV & \,\,\,\, $0.433~$KeV\,\,\,& \,\,\,\, $557.2~$KeV \,\,\,\\
%\hline
%\,\, ~ & \,\,\, ~ \,\,\,& \,\,\, $ 1.5 $ \,\,\,& \,\,\,\,\,\,\,\, $33.02~$KeV & \,\,\,\, $0.435~$KeV\,\,\,& \,\,\,\, $567.3~$KeV \,\,\,\\
%\hline
%\,\, $10^\circ $ & \,\,\, $ 2.7-2.8 $ \,\,\,& \,\,\, $ 0.5 $ \,\,\,& \,\,\,\,\,\,\,\, $26.69 - 27.84$ & \,\,\,\, $0.354 - 0.366$\,\,\,& \,\,\,\, $435.3 - 268.5 $\,\,\,\\
%\hline
%\,\, ~ & \,\,\, ~ \,\,\,& \,\,\, $ 1.0 $ \,\,\,& \,\,\,\,\,\,\,\, $26.94 - 28.06 $ & \,\,\,\, $0.356 - 0.367$\,\,\,& \,\,\,\, $444.3 - 273.6 $\,\,\,\\
%\hline
%\,\, ~ & \,\,\, ~ \,\,\,& \,\,\, $ 1.5 $ \,\,\,& \,\,\,\,\,\,\,\, $27.17 - 28.25$ & \,\,\,\, $0.358 - 0.367$\,\,\,& \,\,\,\, $452.9 - 278.5$\,\,\,\\
%\hline
\,\, ~ & \,\, ~ \,\,\,& \,\,\, $ 0.5 $ \,\,\,& \,\,\,\, $22.41 - 21.97$ & \,\,\,\, $0.294 - 0.276$ \,\,\,& \,\,\,\, $207.5 - 8.335$ \,\,\, \\
%\hline
\,\, $9^\circ$ & \,\,\, $ 2.8 - 3.0$ \,\,\,& \,\,\, $ 1.0 $ \,\,\,& \,\,\,\, $22.60 - 22.40$ & \,\,\,\, $0.296 - 0.281$ \,\,\,& \,\,\,\, $211.8 - 8.618$ \,\,\, \\
%\hline
\,\, ~ & \,\,\, ~ \,\,\,& \,\,\, $ 1.5 $ \,\,\,& \,\,\,\, $22.78 - 22.76$ & \,\,\,\, $0.296 - 0.285$ \,\,\,& \,\,\,\, $215.9 - 8.880$ \,\,\, \\
\hline \hline
\end{tabular}
%\end{center}
\end{table}

%Our numerical results show that, for the strong decays $X \to
%\chi_{cJ}\pi^0$, the $X \to \chi_{c0}\pi^0$ decay is a dominant
%channel and much larger than the channel detected experimentally. We
%also find, the channel $X \to \chi_{c1}\pi^0$ is much smaller than
%the other channels analyzed in this paper so it is difficult to
%observe it if the present model is reasonable. For channel $X \to
%\chi_{c2}\pi^0$, we find it strongly depends on the size parameter
%$\Lambda_{_X}$ which is due to the different $\Lambda_{_X}$ behavior
%and cancelism of the different terms in the coefficient
%$G_{X\chi_{c2}\pi^0}^{(2)}$ so it is difficult to draw any
%conclusion about this channel.

From this Table, one may find that the partial width of decay
$X(3872) \to \chi_{c2}\pi^0$ has a strong dependence on the size
parameter $\Lambda_{_X}$. This can be understood as follows: From
dimensional analysis, in
Eqs.~(\ref{matrixps},\ref{matrixpa},\ref{matrixpt}), only
$G_{X\chi_{c2}\pi^0}^{(2)}$, which gives the dominant contribution
to this process, has a positive mass dimension so it strongly
depends on $\Lambda_{_X}$.

From Table.~\ref{table:width1pi}, one can yield the following ratio
of the partial widths
\begin{eqnarray}
\Gamma(X \to \chi_{c0}\pi^0):\Gamma(X \to \chi_{c1}\pi^0) & \simeq &
1:0.013, \label{eq:ratiotensor}
\end{eqnarray}
%\begin{eqnarray}
%\Gamma(X \to \chi_{c2}\pi^0):\Gamma(X \to \chi_{c1}\pi^0):\Gamma(X
%\to \chi_{c0}\pi^0) & \simeq & 1:0.0027:0.21, \label{eq:ratiotensor}
%\end{eqnarray}
which indicates that, compared with the decay $X \to
\chi_{c0}\pi^0$, $X \to \chi_{c1}\pi^0$ is strongly suppressed. From
the expressions for the partial widths, using the expressions for
the coupling constants in
Eqs.~(\ref{eq:gdstdstpidstdpi},\ref{eq:gchic0dd},\ref{eq:gchic1dstd}),
one naively has the ratio
\begin{eqnarray}
\frac{\Gamma(X \to \chi_{c0}\pi^0)}{\Gamma(X \to \chi_{c1}\pi^0)} &
\simeq & \frac{8}{3
m_{_X}^2}\Big(\frac{4\sqrt{2}}{\sqrt{3}}(m_{D^{\ast}} + 3m_D)\Big)^2
\simeq  \frac{8}{3
m_{_X}^2}\Big(\frac{8\sqrt{2}}{\sqrt{3}}m_{X}\Big)^2 \simeq 110,
\label{eq:estimatewidthratio}
\end{eqnarray}
%\begin{eqnarray}
%\frac{\Gamma(X \to \chi_{c0}\pi^0)}{\Gamma(X \to \chi_{c1}\pi^0)} &
%\simeq & \frac{8}{3 m_{_X}^2}\frac{G_{X \chi_{c0}\pi^0}^2}{G_{X
%\chi_{c1}\pi^0}^2},
%\end{eqnarray}
which can approximately explain the ratio (\ref{eq:ratiotensor}).

%To derive this ratio (\ref{eq:estimatewidthratio}), we did not
%include the contributions from the loop integral and
%$\lambda^{2}(m_{_X}^2,m_{\chi_{c0}}^2,m_{\pi^0}^2)P^{\ast}_{\chi_{c0}}/(\lambda^{2}(m_{_X}^2,m_{\chi_{c1}}^2,m_{\pi^0})P^{\ast}_{\chi_{c1}})$.
%This is reasonable once one takes into account the fact that the
%suppression from the ratio of the loop integral can be compensated
%by the ratio
%$\lambda^{2}(m_{_X}^2,m_{\chi_{c0}}^2,m_{\pi^0}^2)P^{\ast}_{\chi_{c0}}/(\lambda^{2}(m_{_X}^2,m_{\chi_{c1}}^2,m_{\pi^0})P^{\ast}_{\chi_{c1}})$.

Similar to the discussions given in the previous subsection, the
introduction of other components may change our numerical results.
In case of the $c\bar{c}$ component is included, the magnitudes for
the partial widths might be changed but the ratio for the partial
widths must be kept since the $c\bar{c}$ component is a definitely
isospin singlet so it does not contribute to the isospin violating
decays $X(3872) \to \chi_{cJ}\pi^0$. The same conclusion can be
drown if $J/\psi\omega$ are constituents of $X(3872)$. However
complication arises if $X(3872)$ has a $J/\psi\rho$ component as was
done for $J^{PC} = 1^{++}$~\cite{Dong:2009yp}. This is because the
component $J/\psi\rho$ gives a contribution to the isospin violating
$X(3872) \to \chi_{cJ}\pi^0$ decays.

From this, in the case $X(3872)$ with $J^{PC} = 2^{-+}$, the strong
suppression of the decay $X(3872) \to \chi_{c1}\pi^0$ compared with
the decay $X(3872) \to \chi_{c0}\pi^0$ may signal the pure
$DD^{\ast}$ molecular structure of $X(3872)$.

\section{Conclusion}

\label{sec:conclusion}

In this paper, by regarding the hidden charm state $X(3872)$ as a
$DD^{\ast}$ bound state, we studied its radiative and strong decays
in the effective Lagrangian approach. We find, with an approximate
probability of $97.6 \%$ isospin singlet component and a $2.4 \%$
isospin triplet component, the ratios $\mathcal{B}(X\to \gamma
\psi(2S))/\mathcal{B}(X\to \gamma J/\psi)$ by Babar~\cite{:2008rn},
$\mathcal{B}(X\to J/\psi \pi^+\pi^-\pi^0)/\mathcal{B}(X \to
J/\psi\pi^+\pi^-)$ by Belle~\cite{Abe:2005ix} and $\mathcal{B}(X\to
\gamma J/\psi)/\mathcal{B}(X \to J/\psi \pi^+\pi^-)$ by
Babar~\cite{:2008rn} can be explained consistently, but the ratio
$\mathcal{B}(X\to \gamma J/\psi)/\mathcal{B}(X \to J/\psi
\pi^+\pi^-)$ by Belle~\cite{Abe:2005ix} cannot be reproduced. We
would like to stress that, in case of $X(3872)$ as a tensor meson,
few percent of isospin one component of $X(3872)$ in the wave
function can accommodate the large isospin violating partial width
since the phase space of the decay $X(3872) \to J/\psi
\pi^+\pi^-\pi^0$ is about one percent of that of the decay $X(3872)
\to J/\psi \pi^+\pi^-$.

With respect to the data, we fixed both of the parameters $\phi$ and
$\Lambda_{_X}$ in our model, so that all the partial widths can be
calculated explicitly. We calculated the strong decays $X(3872) \to
J/\psi \pi^+\pi^-\pi^0$, $X(3872) \to J/\psi \pi^+\pi^-$, $X(3872)
\to \chi_{cJ}\pi^0~(J = 0,1,2)$ and radiative decays $X(3872) \to
\gamma J/\psi, \gamma\psi(2S)$. For the strong decays $X(3872) \to
J/\psi\pi^+\pi^-$ and $X(3872) \to J/\psi\pi^+\pi^-\pi^0$, our
results are several KeV which are both around one order smaller than
the case $X(3872)$ with $J^{PC} = 1^{++}$. For radiative decays, our
results are both of order KeV. They are at the same order as those
in the case that $X(3872)$ with $J^{PC} = 1^{++}$ is a bound state
of mesons~\cite{Dong:2008gb}, while much smaller than those obtained
in the case that $X(3872)$ is a mixing state of molecule and
charmonium component~\cite{Dong:2009yp}.

For the $X(3872) \to \chi_{cJ}\pi^0$ decays, we found that, compared
with the decay $X(3872) \to \chi_{c0}\pi^0$, the $X(3872) \to
\chi_{c1}\pi^0$ decay is strongly suppressed. The experimental
observation of this suppression may be a signal of the pure
$DD^{\ast}$ molecular structure of $X(3872)$.

Concerning the lessons from the study of $X(3872)$ with $J^{PC} =
1^{++}$, we naively expect that the inclusion of other constituent
in the wave function may change our numerical results of the
relevant partial widths. Explicitly, the inclusion of $J/\psi\omega$
or/and $J/\psi\rho$ components may increase some partial widths for
the strong decays while the inclusion of $c\bar{c}$ component may
increase the partial width for radiative decay. As was noted, our
above calculation cannot yield the ratio $\mathcal{B}(X\to \gamma
J/\psi)/\mathcal{B}(X \to J/\psi \pi^+\pi^-)$ by
Belle~\cite{Abe:2005ix} although we reproduced this ratio by
Babar~\cite{:2008rn}. If the Belle value is preferred in the future
experiment, in the molecular interpretation, we should include other
constituents. We leave this inclusion of other constituents in our
future work.

Taking into the quantum numbers of $X(3872)$ and its constituents,
the coupling between $X(3872)$ and $DD^{\ast}$ is via $P-$wave. We
would like to mention that, although the meson exchanging model
calculation indicates that it is difficult to form a $DD^{\ast}$
molecule via $P-$wave coupling~\cite{Liu:2008tn}, we think that it
deserves systematically and seriously analysis of the exchanged
hadrons and coupling channel effects. In the present analysis, we do
not consider origin of the binding force but regard $X(3872)$ as a
$P-$wave bound state of $DD^{\ast}$.

In conclusion, by regarding that the hidden charm meson $X(3872)$
with $J^{PC} = 2^{-+}$ is a $DD^{\ast}$ $P-$wave bound state, we
found that our model with dominant isospin zero component can
explain the existing date quite well. We also predicted the partial
widths for $X(3872) \to \gamma J/\psi$, $X(3872) \to
\gamma\psi(2S)$, $X(3872) \to J/\psi \pi^+\pi^-$, $X(3872) \to
J/\psi \pi^+\pi^-\pi^0$ and $X(3872) \to \chi_{cJ}\pi^0$. Comparison
of these values with the future experiments will shed a light on the
nature of $X(3872)$.

%%%%%%%%%%%%%%%%%%%%%%%%%%%%%%%%%%%%%%%%%%%%%%%%%%%%%%%%%%%%%%%%%%%
\appendix

\section{Explicit forms for the functions $C_{X\psi_n\gamma}^{ij}$ for radiative decays.}

\label{app:cijrad}

Here, we present the nonvanishing coefficients $C_{X\psi_n
\gamma}^{ij}$ for the radiative decay of $X(3872) \to \gamma \psi_n$
with $\psi_1 = J/\psi$ and $\psi_2 = \psi(2S)$:
\begin{eqnarray}
C_{X\psi_n \gamma}^{11} & = & \frac{1}{3m_X^2
m_{\psi_n}^2}\lambda(3\lambda + 40 m_{_X}^2 m_{\psi_n}^2) ,
\;\;\;\;\;\;\;\;\;\;\;\;\;\;\;\;
%%%%%%%%%%%%%%%%%%%%%%%%%%%%%%%%%%%%%%%%%%%%%%%%%%%%%%%%%%%%%%%%%%%%%%%%%%%%%%%%%%%
C_{X\psi_n \gamma}^{12} = - \frac{40}{3} \lambda , \nonumber\\
%%%%%%%%%%%%%%%%%%%%%%%%%%%%%%%%%%%%%%%%%%%%%%%%%%%%%%%%%%%%%%%%%%%%%%%%%%%%%%%%%%%
C_{X\psi_n \gamma}^{13} & = & \frac{4}{3m_X^2 m_{\psi_n}^2}
\lambda(3\lambda + 10 m_{_X}^2 m_{\psi_n}^2 - 10 m_{\psi_n}^4) ,
\;\;\;\;
%%%%%%%%%%%%%%%%%%%%%%%%%%%%%%%%%%%%%%%%%%%%%%%%%%%%%%%%%%%%%%%%%%%%%%%%%%%%%%%%%%%
C_{X\psi_n \gamma}^{14} = - \frac{2}{m_X^2 m_{\psi_n}^2}
(m_{_X}^2 + m_{\psi_n}^2)\lambda^2 , \nonumber\\
%%%%%%%%%%%%%%%%%%%%%%%%%%%%%%%%%%%%%%%%%%%%%%%%%%%%%%%%%%%%%%%%%%%%%%%%%%%%%%%%%%%
C_{X\psi_n \gamma}^{15} & = & \frac{8}{3m_X^2} \lambda^2 ,
\;\;\;\;\;\;\;\;
%%%%%%%%%%%%%%%%%%%%%%%%%%%%%%%%%%%%%%%%%%%%%%%%%%%%%%%%%%%%%%%%%%%%%%%%%%%%%%%%%%%
C_{X\psi_n \gamma}^{22} = \frac{28}{3}\lambda , \;\;\;\;\;\;\;\;
%%%%%%%%%%%%%%%%%%%%%%%%%%%%%%%%%%%%%%%%%%%%%%%%%%%%%%%%%%%%%%%%%%%%%%%%%%%%%%%%%%%
C_{X\psi_n \gamma}^{23} = \frac{16}{3m_{_X}^2}(m_{_X}^2 -
m_{\psi_n}^2)\lambda , \;\;\;\;\;\;\;\;
%%%%%%%%%%%%%%%%%%%%%%%%%%%%%%%%%%%%%%%%%%%%%%%%%%%%%%%%%%%%%%%%%%%%%%%%%%%%%%%%%%%
C_{X\psi_n \gamma}^{25} = \frac{8}{3m_{_X}^2}\lambda^2 , \nonumber\\
%%%%%%%%%%%%%%%%%%%%%%%%%%%%%%%%%%%%%%%%%%%%%%%%%%%%%%%%%%%%%%%%%%%%%%%%%%%%%%%%%%%
C_{X\psi_n \gamma}^{33} & = & \frac{4}{3m_X^4m_{\psi_n}^2}(3m_{_X}^2
+ 4m_{\psi_n}^2) \lambda^2 , \;\;\;\;\;\;\;\;
%%%%%%%%%%%%%%%%%%%%%%%%%%%%%%%%%%%%%%%%%%%%%%%%%%%%%%%%%%%%%%%%%%%%%%%%%%%%%%%%%%%
C_{X\psi_n \gamma}^{34} = - \frac{4}{m_X^2m_{\psi_n}^2}(m_{_X}^2 -
m_{\psi_n}^2) \lambda^2 , \nonumber\\
%%%%%%%%%%%%%%%%%%%%%%%%%%%%%%%%%%%%%%%%%%%%%%%%%%%%%%%%%%%%%%%%%%%%%%%%%%%%%%%%%%%
C_{X\psi_n \gamma}^{35} & = & \frac{16}{3m_X^4}(m_{_X}^2 -
m_{\psi_n}^2) \lambda^2 , \;\;\;\;\;\;\;\;
%%%%%%%%%%%%%%%%%%%%%%%%%%%%%%%%%%%%%%%%%%%%%%%%%%%%%%%%%%%%%%%%%%%%%%%%%%%%%%%%%%%
C_{X\psi_n \gamma}^{44} = \frac{1}{m_X^2m_{\psi_n}^2} \lambda^3 ,
\;\;\;\;\;\;\;\;
%%%%%%%%%%%%%%%%%%%%%%%%%%%%%%%%%%%%%%%%%%%%%%%%%%%%%%%%%%%%%%%%%%%%%%%%%%%%%%%%%%%
C_{X\psi_n \gamma}^{55} = \frac{4}{3m_{_X}^4} \lambda^3 ,
\end{eqnarray}
where $\lambda = (m_{_X}^2 - m_{\psi_n}^2)^2$.

\section{Explicit forms for the functions $C^{ij}$ for strong decays.}

\subsection{Explicit forms for the functions $C_{XJ/\psi V}^{ij}$ for $X(3872) \to J/\psi h$
decays.}

\label{app:cijstr1}

In this subsection, we list the nonvanishing coefficients
$C_{XJ/\psi V}^{ij}$ for the strong decay of $X(3872) \to J/\psi h$:
\begin{eqnarray}
C_{XJ/\psi V}^{11} & = &
\frac{1}{3m_{_X}^2m_{J/\psi}^2}\lambda(3\lambda + 40m_{_X}^2
m_{J/\psi}^2) , \;\;\;\;\;\;\;\;\;\;\;\;\;\;\; C_{XJ/\psi V}^{12} = -\frac{40}{3}\lambda , \nonumber\\
%%%%%%%%%%%%%%%%%%%%%%%%%%%%%%%%%%%%%%%%%%%%%%%%%%%%%%%%%%%%%%%%%%%%%%%%%%%%%%%%%%%
C_{XJ/\psi V}^{13} & = & \frac{2}{3m_{_X}^2m_{J/\psi}^2}\lambda(3\lambda + 20m_{_X}^2 m_{J/\psi}^2), \nonumber\\
%%%%%%%%%%%%%%%%%%%%%%%%%%%%%%%%%%%%%%%%%%%%%%%%%%%%%%%%%%%%%%%%%%%%%%%%%%%%%%%%%%%
C_{XJ/\psi V}^{14} & = & \frac{2}{3m_{_X}^2m_{J/\psi}^2}\lambda(3\lambda + 20m_{_V}^2 m_{J/\psi}^2 - 20m_{J/\psi}^4) , \nonumber\\
%%%%%%%%%%%%%%%%%%%%%%%%%%%%%%%%%%%%%%%%%%%%%%%%%%%%%%%%%%%%%%%%%%%%%%%%%%%%%%%%%%%
C_{XJ/\psi V}^{15} & = &
-\frac{2}{m_{_X}^2m_{J/\psi}^2}\lambda^2(m_{_X}^2 - m_{_V}^2 +
m_{J/\psi}^2 ) , \;\;\;\;\;\;\;\;
%%%%%%%%%%%%%%%%%%%%%%%%%%%%%%%%%%%%%%%%%%%%%%%%%%%%%%%%%%%%%%%%%%%%%%%%%%%%%%%%%%%
C_{XJ/\psi V}^{17} = \frac{8}{3m_{_X}^2}\lambda^2 , \nonumber\\
%%%%%%%%%%%%%%%%%%%%%%%%%%%%%%%%%%%%%%%%%%%%%%%%%%%%%%%%%%%%%%%%%%%%%%%%%%%%%%%%%%%
C_{XJ/\psi V}^{22} & = & \frac{1}{3m_{_X}^2m_{_V}^2}\lambda(3\lambda
+ 40m_{_X}^2 m_{_V}^2) , \;\;\;\;\;\;\;\;\;\;\;\;\;\;\;\;\;\;\;\;\;
%%%%%%%%%%%%%%%%%%%%%%%%%%%%%%%%%%%%%%%%%%%%%%%%%%%%%%%%%%%%%%%%%%%%%%%%%%%%%%%%%%%
C_{XJ/\psi V}^{23} = \frac{2}{3m_{_X}^2m_{_V}^2}\lambda(3\lambda + 20m_{_X}^2 m_{_V}^2) , \nonumber\\
C_{XJ/\psi V}^{24} & = & - \frac{2}{3m_{_X}^2m_{_V}^2}\lambda(3\lambda + 20m_{_V}^2 m_{J/\psi}^2 - 20m_{_V}^4) , \nonumber\\
%%%%%%%%%%%%%%%%%%%%%%%%%%%%%%%%%%%%%%%%%%%%%%%%%%%%%%%%%%%%%%%%%%%%%%%%%%%%%%%%%%%
C_{XJ/\psi V}^{26} & = &
-\frac{2}{m_{_X}^2m_{_V}^2}\lambda^2(m_{_X}^2 + m_{_V}^2 -
m_{J/\psi}^2 ) , \;\;\; \;\;\; \;\;\;
%%%%%%%%%%%%%%%%%%%%%%%%%%%%%%%%%%%%%%%%%%%%%%%%%%%%%%%%%%%%%%%%%%%%%%%%%%%%%%%%%%%
C_{XJ/\psi V}^{27} = \frac{8}{3m_{_X}^2}\lambda^2 , \nonumber\\
%%%%%%%%%%%%%%%%%%%%%%%%%%%%%%%%%%%%%%%%%%%%%%%%%%%%%%%%%%%%%%%%%%%%%%%%%%%%%%%%%%%
C_{XJ/\psi V}^{33} & = & \frac{1}{3m_{_X}^2m_{_V}^2m_{J/\psi}^2}\lambda\Big[3(m_{V}^2 + m_{J/\psi}^2)\lambda + 40m_{_X}^2 m_{_V}^2m_{J/\psi}^2 \Big] , \nonumber\\
%%%%%%%%%%%%%%%%%%%%%%%%%%%%%%%%%%%%%%%%%%%%%%%%%%%%%%%%%%%%%%%%%%%%%%%%%%%%%%%%%%%
C_{XJ/\psi V}^{34} & = & \frac{2}{3m_{_X}^2m_{_V}^2m_{J/\psi}^2}(m_{_V}^2-m_{J/\psi}^2)\lambda\Big[3\lambda + 40 m_{_V}^2m_{J/\psi}^2 \Big] , \nonumber\\
%%%%%%%%%%%%%%%%%%%%%%%%%%%%%%%%%%%%%%%%%%%%%%%%%%%%%%%%%%%%%%%%%%%%%%%%%%%%%%%%%%%
C_{XJ/\psi V}^{35} & = & - \frac{2}{m_{_X}^2m_{J/\psi}^2}(m_{_X}^2 - m_{_V}^2 + m_{J/\psi}^2)\lambda^2 , \nonumber\\
%%%%%%%%%%%%%%%%%%%%%%%%%%%%%%%%%%%%%%%%%%%%%%%%%%%%%%%%%%%%%%%%%%%%%%%%%%%%%%%%%%%
C_{XJ/\psi V}^{36} & = & - \frac{2}{m_{_X}^2m_{V}^2}(m_{_X}^2 +
m_{_V}^2 - m_{J/\psi}^2)\lambda^2 ,  \;\;\; \;\;\; \;\;\; \;
%%%%%%%%%%%%%%%%%%%%%%%%%%%%%%%%%%%%%%%%%%%%%%%%%%%%%%%%%%%%%%%%%%%%%%%%%%%%%%%%%%%
C_{XJ/\psi V}^{37} = \frac{16}{3m_{_X}^2}\lambda^2 , \nonumber\\
%%%%%%%%%%%%%%%%%%%%%%%%%%%%%%%%%%%%%%%%%%%%%%%%%%%%%%%%%%%%%%%%%%%%%%%%%%%%%%%%%%%
C_{XJ/\psi V}^{44} & = & \frac{1}{3m_{_X}^4 m_{_V}^2 m_{J/\psi}^2}\lambda\Big[3m_{_X}^2(m_{_V}^2 + m_{J/\psi}^2)\lambda - 24 m_{_V}^2 m_{J/\psi}^2 \lambda + 40 m_{_V}^2 m_{J/\psi}^2 (m_{_V}^2 - m_{J/\psi}^2)^2\Big] , \nonumber\\
%%%%%%%%%%%%%%%%%%%%%%%%%%%%%%%%%%%%%%%%%%%%%%%%%%%%%%%%%%%%%%%%%%%%%%%%%%%%%%%%%%%
C_{XJ/\psi V}^{45} & = & - \frac{2}{m_{_X}^2 m_{J/\psi}^2}(m_{_X}^2
- m_{_V}^2 - 3m_{J/\psi}^2)\lambda^2 ,  \;\;\; \;\;\; \;\;\; \;\;\;
%%%%%%%%%%%%%%%%%%%%%%%%%%%%%%%%%%%%%%%%%%%%%%%%%%%%%%%%%%%%%%%%%%%%%%%%%%%%%%%%%%%
C_{XJ/\psi V}^{46} = \frac{2}{m_{_X}^2 m_{_V}^2}(m_{_X}^2 - 3m_{_V}^2 - m_{J/\psi}^2)\lambda^2 , \nonumber\\
%%%%%%%%%%%%%%%%%%%%%%%%%%%%%%%%%%%%%%%%%%%%%%%%%%%%%%%%%%%%%%%%%%%%%%%%%%%%%%%%%%%
C_{XJ/\psi V}^{47} & = & \frac{16}{3m_{_X}^4 }(m_{_V}^2 -
m_{J/\psi}^2)\lambda^2 ,  \;\;\;
%%%%%%%%%%%%%%%%%%%%%%%%%%%%%%%%%%%%%%%%%%%%%%%%%%%%%%%%%%%%%%%%%%%%%%%%%%%%%%%%%%%
C_{XJ/\psi V}^{55} = \frac{1}{m_{_X}^2 m_{J/\psi}^2 }\lambda^3 ,
\;\;\;
%%%%%%%%%%%%%%%%%%%%%%%%%%%%%%%%%%%%%%%%%%%%%%%%%%%%%%%%%%%%%%%%%%%%%%%%%%%%%%%%%%%
C_{XJ/\psi V}^{66} = \frac{1}{m_{_X}^2 m_{_V}^2 }\lambda^3 ,  \;\;\;
%%%%%%%%%%%%%%%%%%%%%%%%%%%%%%%%%%%%%%%%%%%%%%%%%%%%%%%%%%%%%%%%%%%%%%%%%%%%%%%%%%%
C_{XJ/\psi V}^{77} = \frac{4}{3m_{_X}^4}\lambda^3 ,
\end{eqnarray}
where $\lambda = \lambda(m_{X}^2,m_{J/\psi}^2,q_1^2) =
\Big(m_{X}^2-(m_{J/\psi}-q_1)^2\Big)\Big(m_{X}^2-(m_{J/\psi}+q_1)^2\Big)$
is the K$\ddot{a}$llen function with $q_1 = \sqrt{q_1^2}$.

\subsection{Explicit forms for the functions $C^{ij}$ for $X(3872) \to \chi_{c2}\pi^0$
decay.}

\label{app:cijstr2}

Here, we present the nonvanishing coefficients
$C_{X\chi_{c2}\pi^0}^{ij}$ for the strong decay of $X(3872) \to
\chi_{c2}\pi^0$:
\begin{eqnarray}
C_{X\chi_{c2}\pi^0}^{11} & = & \frac{1}{576m_X^4
m_{\chi_{c2}}^4}\lambda^4 , \;\;\;\;\;\;\;\;
%%%%%%%%%%%%%%%%%%%%%%%%%%%%%%%%%%%%%%%%%%%%%%%%%%%%%%%%%%%%%%%%%%%%%%%%%%%%%%%%%%%
C_{X\chi_{c2}\pi^0}^{12} = \frac{1}{36m_X^4 m_{\chi_{c2}}^4}
\lambda^2(\lambda + 6m_X^2m_{\chi_{c2}}^2) , \nonumber\\
%%%%%%%%%%%%%%%%%%%%%%%%%%%%%%%%%%%%%%%%%%%%%%%%%%%%%%%%%%%%%%%%%%%%%%%%%%%%%%%%%%%
C_{X\chi_{c2}\pi^0}^{13} & = & \frac{1}{36m_X^4 m_{\chi_{c2}}^4}
(m_X^2+m_{\chi_{c2}}^2-m_{\pi^0}^2)\lambda^3 , \nonumber\\
%%%%%%%%%%%%%%%%%%%%%%%%%%%%%%%%%%%%%%%%%%%%%%%%%%%%%%%%%%%%%%%%%%%%%%%%%%%%%%%%%%%
C_{X\chi_{c2}\pi^0}^{22} & = & \frac{1}{9m_X^4 m_{\chi_{c2}}^4}
\Big[\lambda^2 + 30 m_X^2 m_{\chi_{c2}}^2\lambda + 180 m_X^4 m_{\chi_{c2}}^4\Big] , \nonumber\\
%%%%%%%%%%%%%%%%%%%%%%%%%%%%%%%%%%%%%%%%%%%%%%%%%%%%%%%%%%%%%%%%%%%%%%%%%%%%%%%%%%%
C_{X\chi_{c2}\pi^0}^{23} & = & \frac{2}{9m_X^4 m_{\chi_{c2}}^4}
(m_X^2+m_{\chi_{c2}}^2-m_{\pi^0}^2)\lambda(\lambda + 15 m_X^2 m_{\chi_{c2}}^2) , \nonumber\\
%%%%%%%%%%%%%%%%%%%%%%%%%%%%%%%%%%%%%%%%%%%%%%%%%%%%%%%%%%%%%%%%%%%%%%%%%%%%%%%%%%%
C_{X\chi_{c2}\pi^0}^{33} & = & \frac{1}{18m_X^4 m_{\chi_{c2}}^4}
\lambda^2(2\lambda + 17 m_X^2 m_{\chi_{c2}}^2),
\end{eqnarray}
where $\lambda =
\lambda(m_{X}^2,m_{\chi_{c2}}^2,m_{\pi^0}^2)=\Big(m_{X}^2-(m_{\chi_{c2}}-m_{\pi^0})^2\Big)\Big(m_{X}^2-(m_{\chi_{c2}}+m_{\pi^0})^2\Big)$
is the K$\ddot{a}$llen function.

\section{Decay widths for $\omega \to \pi^+\pi^-\pi^0$.}

\label{app:omega3pi}

From the explicit study of Ref.~\cite{Braaten:2005ai} we have the
following expression for the $\omega \to \pi^+\pi^-\pi^0$ decay that
is used in this work.
\begin{eqnarray}
\int d\Pi  F(\omega \to \pi^+ \pi^- \pi^0) & = &\frac{(\cos\theta_v
+ \sqrt{2} \sin\theta_v)^2}{32\pi^3F_\pi^6q_1^2}
\int_{4m_\pi^2}^{(q_1-m_\pi)^2}ds_{12}\int_{s_{23}^-}^{s_{23}^+}ds_{23}
\Big[ m_\pi^2(m_\pi^2-q_1^2)^2 - s_{12}s_{23}s_{13}\Big]\nonumber\\
& & \times \Big|\Big[C_{v3\pi} +
\frac{2g_{\rho\pi\pi}C_{vv\pi}F_\pi^2}{m_v^2}\Big(1 -
\frac{1}{3}[f_\rho(s_{12}) +
f_\rho(s_{23})+f_\rho(s_{31})]\Big)\Big]  \Big|^2,
\end{eqnarray}
where $s_{12}, s_{23}$ and $s_{13}$ are the invariant mass squares
of the three pion pairs and
\begin{eqnarray}
s_{23}^{\pm} & = & m_\pi^2 + \frac{1}{2}\Big(q_1^2 + m_\pi^2 -
s_{12} \pm
\lambda^{1/2}(s_{12},q_1^2,m_\pi^2)\sqrt{1-\frac{4m_\pi^2}{s_{12}}}\big),
\end{eqnarray}
and
\begin{eqnarray}
f_{V}(s)\equiv {s \over s-M_{V}^2+im_{V}\Gamma_{V}},
\end{eqnarray}
is a vector meson resonance factor that vanishes at $s=0$. The pion
decay constant is $F_{\pi}= 92.4$ MeV, the values of the parameters
$C_{v 3\pi}$ and $g_{\rho \pi\pi}C_{vv\pi}F_{\pi}^2/m_v^2$ are given
by Eqs.~(\ref{Cv3pi}) and (\ref{GCvvpi}), and the value of the light
vector meson mixing angle $\theta_{v}$ is given by
Eq.~(\ref{costheta})~\cite{Braaten:2005ai}
\begin{eqnarray}
C_{v 3\pi}+2g_{\rho \pi\pi} C_{vv\pi} F_{\pi}^2/m_v^2 & = & (8.03
\pm
0.48)/(16\pi^2), \label{Cv3pi} \\
g_{\rho \pi\pi} C_{vv\pi} F_{\pi}^2/m_v^2 & = & (5.1 \pm
0.65)/(16\pi^2).\label{GCvvpi}
\end{eqnarray}
The vector meson mixing angle is given
by~\cite{Braaten:1989zn,Braaten:1987jh}
\begin{equation}
\cos \theta_v = 0.51 \pm 0.01. \label{costheta}
\end{equation}
The function $\cos\theta_v+\sqrt{2}\sin\theta_v$ used in the
numerical function is $\cos\theta_v+\sqrt{2}\sin\theta_v \approx
1.73\pm 0.01$. The errors in the parameters in Eqs.~(\ref{Gvpp}),
(\ref{Cv3pi}), (\ref{GCvvpi}), and (\ref{costheta}) are determined
using the uncertainties in the measurements of the vector meson
decay widths only.

%%%%%%%%%%%%%%%%%%%%%%%%%%%%%%%%%%%%%%%%%%%%%%%%%%%%%%%%%%%%%%%%%%%%%%%
\acknowledgments

\label{ACK}

This work is supported in part by Grant-in-Aid for Scientific
Research on Innovative Areas (No. 2104) ``Quest on New Hadrons with
Variety of Flavors'' from MEXT. The work of M.H. is supported in
part by the Grant-in-Aid for Nagoya University Global COE Program
``Quest for Fundamental Principles in the Universe: from Particles
to the Solar System and the Cosmos'' from MEXT, the JSPS
Grant-in-Aid for Scientific Research (S) $\sharp$ 22224003, (c)
$\sharp$ 20540262. The work of Y.M. is supported in part by the
National Science Foundation of China (NNSFC) under grant No.
10905060.

%%%%%%%%%%%%%%%%%%%%%%%%%%%%%%%%%%%%%%%%%%%%%%%%%%%%%%%%%%%%%%%%%%%%%%%%%%%%%%%%%%

\end{document}